# The locality of quantum subsystems II: A 'local sum-over-paths' ontology for quantum subsystems

Adam Brownstein[1]

[1]*University of Melbourne, School of Physics*

23/08/2017

Subsystems of entangled quantum systems are not traditionally described in a local way. This paper begins to address the issue by constructing an explicit local hidden variable theory for quantum subsystems. The interpretation is based on a modified sum-over-paths approach, where the paths of the subsystem obtain information about the external system during interactions. The interpretation utilizes the conception that there is a dissociation between the tensor product structure of wavefunctions and information flow into quantum subsystems.

# 1 Introduction

Describing the physics of subsystems is central to quantum theory. Although quantum mechanics in principle predicts the measurement outcomes of the total system, all observed phenomena consist of the measurement results of quantum subsystems in practice. In addition, it is plausible that subsystems play an important role in the ontology of the total quantum system itself. These two aspects lead to the question of whether quantum subsystems can be described independently to the total system. Paper one of this series [1] has demonstrated that the standard ontologies of quantum mechanics do not contain a local description of quantum subsystems, and so the question is open to further exploration. Here, we continue our investigation into the physical nature of quantum subsystems and begin to address this issue.

We establish in this paper a general framework for the local interpretation of marginal probabilities in quantum mechanics. Our presentation begins in the two particle case, where a *local sum-over-paths* approach is used to produce the marginal probabilities of a single-particle subsystem. In this local sum-over-paths interpretation, it is seen that interactions with particles of the environment impart hits of information onto the paths of the subsystem, where these these hits of information are used to recursively update local hidden variables. The hidden variables control the interference terms which give quantum mechanical deviations from a classical sum over transition probabilities.

The methods developed in the two particle case are then generalized to the three-particle case. The three-particle case is seen to have an additional layer of ontological complexity. However, the local sum-over-paths interpretation remains valid. It becomes clear that this method can be generalized to the N-particle case, and so is a suitable way to describe quantum subsystems.

# 2 A local sum-over-paths interpretation for the quantum subsystem

This section establishes the local sum-over-paths interpretation in the two particle case. Section 2.1 gives some preliminary information about the standard sum-over-paths approach to quantum mechanics. Sections 2.2 and 2.3 then present the local sum-over-paths approach. We use the local sum-over-paths interpretation to explain the marginals of an EPR-B experiment.

## 2.1 Converting unitary matrix notation into a sum-over-paths

Before we begin our presentation of the local sum-over-paths interpretation, we first establish the relation between the matrix notation of a quantum circuit and the traditional sum-over-paths representation of it. The first fact required is that any unitary operator can be decomposed into a set of general single-particle gates and controlled phase gates due to universality of this particular gate set. Therefore a general two qubit unitary operator can be expressed in the following way:

$$\mathbb{U}^{(n)} = \mathbb{A}^{(n)} \otimes \mathbb{B}^{(n)} \mathbb{A}\mathbb{B}^{(n)} ... \mathbb{A}^{(2)} \otimes \mathbb{B}^{(2)} \mathbb{A}\mathbb{B}^{(2)} \mathbb{A}^{(1)} \otimes \mathbb{B}^{(1)} \mathbb{A}\mathbb{B}^{(1)}. \qquad (2.1)$$

where $\mathbb{A}^{(t)}$ and $\mathbb{B}^{(t)}$ are single particle gates and $\mathbb{A}\mathbb{B}^{(t)} = \text{diag}\left(e^{i\theta_1^{(t)}}, e^{i\theta_2^{(t)}}, e^{i\theta_3^{(t)}}, e^{i\theta_4^{(t)}}\right)$ is a two qubit controlled phase gate. These matrices are all labeled by a superscript $t$, which denotes that the gate is applied at layer $t$ in the quantum circuit described by $\mathbb{U}^{(n)}$. Similarly, the phase parameters $\theta_s^{(t)}$ of $\mathbb{A}\mathbb{B}^{(t)}$ carry this superscript $t$ to identify them with their corresponding controlled phase gate.

A related fact is that a universal set of gates can produce any quantum state. Therefore it is always possible to describe a quantum state as being prepared from the initial state $|0,0\rangle$ by a unitary matrix decomposed in the manner above. These facts taken together imply that the probability amplitudes of a general two qubit circuit, from state preparation to measurement, can be described by:

$$\langle j,k|\mathbb{U}^{(n)}|0,0\rangle = \langle j,k|\mathbb{A}^{(n)} \otimes \mathbb{B}^{(n)} \mathbb{A}\mathbb{B}^{(n)} ... \mathbb{A}^{(2)} \otimes \mathbb{B}^{(2)} \mathbb{A}\mathbb{B}^{(2)} \mathbb{A}^{(1)} \otimes \mathbb{B}^{(1)} \mathbb{A}\mathbb{B}^{(1)}|0,0\rangle. \qquad (2.2)$$

To convert this amplitude to the sum-over-paths representation, the unitary matrices can be explicitly multiplied out, producing a sum over products of matrix elements. These products of matrix elements are probability amplitudes for particular particle paths. As an example, the probability amplitudes for the state $\mathbb{A}^{(2)} \otimes \mathbb{B}^{(2)} \mathbb{A}\mathbb{B}^{(2)} \mathbb{A}^{(1)} \otimes \mathbb{B}^{(1)} \mathbb{A}\mathbb{B}^{(1)}|0,0\rangle$ to be detected in the modes $|j,k\rangle$ can be described by a sum of four path amplitudes:

$$\langle j,k| \mathbb{A}^{(2)} \otimes \mathbb{B}^{(2)} \mathbb{A}\mathbb{B}^{(2)} \mathbb{A}^{(1)} \otimes \mathbb{B}^{(1)} \mathbb{A}\mathbb{B}^{(1)}|0,0\rangle$$
$$= A^{(2)}_{j0} A^{(1)}_{00} B^{(2)}_{k0} B^{(1)}_{00} e^{i\theta^{(2)}_1} e^{i\theta^{(1)}_1} + A^{(2)}_{j0} A^{(1)}_{00} B^{(2)}_{k1} B^{(1)}_{10} e^{i\theta^{(2)}_2} e^{i\theta^{(1)}_1} \qquad (2.3)$$
$$+ A^{(2)}_{j1} A^{(1)}_{10} B^{(2)}_{k0} B^{(1)}_{00} e^{i\theta^{(2)}_3} e^{i\theta^{(1)}_1} + A^{(2)}_{j1} A^{(1)}_{10} B^{(2)}_{k1} B^{(1)}_{10} e^{i\theta^{(2)}_4} e^{i\theta^{(1)}_1}.$$

Similarly, the probability amplitudes for a general two qubit circuit can be represented in a more condensed notation as a sum over path amplitudes:

$$\langle j,k| \mathbb{U}^{(n)}|0\rangle = \sum_{P(j,n)} \sum_{M(k,n)} A_P B_M e^{i\alpha(P,M)}. \qquad (2.4)$$

The symbol $A_P$ denotes the probability amplitude for path $P$ of particle A, while $B_M$ denotes the probability amplitude of path $M$ of particle B. The paths in the summations $\sum_{P(j,n)}$ and $\sum_{M(k,n)}$ carry indices $(j,n)$ and $(k,n)$ respectively. These indices specify that the paths of particle A to be summed end at mode j at layer n of the circuit, while the paths of particle B to be summed end at mode k at layer n of the circuit. The phase $e^{i\alpha(P,M)}$ is acquired due to the action of controlled phase gates, and affects the amplitude $A_P B_M e^{i\alpha(P,M)}$ for the configuration space path $(P,M)$.

## 2.2 Decomposing the marginal probabilities

To obtain a local sum-over-paths interpretation, the marginal probability is first expressed in a traditional sum-over-paths representation, and then the paths of particle A are singled out in the expression. Let $P(A = j, B = k)$ be the joint probability for particles A and B to be detected in modes $j$ and $k$ respectively. As described in Section 2.1, $P(A = j, B = k)$ is expressible as a sum over configuration space paths as follows:

$$P(A = j, B = k) = |\sum_{P(j,n)} \sum_{M(k,n)} A_P B_M e^{i\alpha(P,M)}|^2. \qquad (2.5)$$

Now let $P(A = j)$ denote the marginal probability of particle A to be detected in mode $j$. $P(A = j)$ is given by a classical sum over the joint probabilities, and by using Eq. (2.5) can be written as:

$$P(A = j) = \sum_k P(A = j, B = k)$$
$$= \sum_k |\sum_{P(j,n)} \sum_{M(k,n)} A_P B_M e^{i\alpha(P,M)}|^2$$
$$= \sum_k |\sum_{P(j,n)} A_P \left( \sum_{M(k,n)} B_M e^{i\alpha(P,M)} \right)|^2$$
$$= \sum_k |\sum_{P(j,n)} A_P \langle k|\mathbb{B}^{(n)}_P|0\rangle|^2, \qquad (2.6)$$

where we have recognized that $\sum_{M(k,n)} B_M e^{i\alpha(P,M)} = \langle k|\mathbb{B}^{(n)}_P|0\rangle$, where $\mathbb{B}^{(n)}_P$ is defined as the unitary matrix giving the transition amplitudes of particle B up to time step n, given that particle A traverses path $P$. The matrix $\mathbb{B}^{(n)}_P$ is distinguished from the matrix $\mathbb{B}^{(n)}$ by the subscript $P$. These two matrices are related, for $\mathbb{B}^{(n)}_P = \mathbb{A}\mathbb{B}^{(n)}_P \mathbb{B}^{(n)} \mathbb{B}^{(n-1)} \mathbb{A}\mathbb{B}^{(n-1)}_P ... \mathbb{A}\mathbb{B}^{(1)}_P \mathbb{B}^{(1)}$. In this expression, the matrix $\mathbb{A}\mathbb{B}^{(t)}_P =$

diag$\{e^{i\theta_0^{(t)}(P)}, e^{i\theta_1^{(t)}(P)}\}$ represents the controlled phase gate applied to particle B at layer $t$ given that particle A traverses path $P$. The phases $\theta_s^{(t)}(P)$ of this controlled phase gate carry the superscript $t$ denoting the layer at they are applied, a subscript $s$ denoting the state to they are applied, and a label $P$ which specifies that the application of the phase is conditional on path $P$ of particle A. We have defined the phases in a condensed notation using $e^{-i\theta_0^{(t)}(P)} \equiv \bar{\eta}^{(k,t)}$, $e^{i\theta_0^{(t)}(Q)} \equiv \eta^{(k,t)}$ which will simplify calculations. Overbars on the symbols $\bar{\eta}^{(k,t)}$ are supposed to indicate complex conjugates of phases, such as $e^{-i\theta_0^{(t)}(P)}$. The overbarred symbols are always associated with paths $P$, while the symbols lacking the overbar, for instance $\eta^{(k,t)}$, are associated with paths $Q$ in the summations. Note that $\mathbb{AB}_P^{(t)}$ is a single-particle gate, and is distinguished from $\mathbb{AB}^{(t)}$ which is a two-particle gate by the subscript $P$. These two gates are related in the following manner:

$$\mathbb{AB}^{(t)} = |0\rangle\langle 0| \otimes \mathbb{AB}_{P_0}^{(t)} + |1\rangle\langle 1| \otimes \mathbb{AB}_{P_1}^{(t)}, \tag{2.7}$$

where we have used $P_0$ to denote any path of particle A which passes through mode $|0\rangle_A$ at this layer, and $P_1$ to denote any path of particle A which passes mode $|1\rangle_A$ at this layer. We want to split the squared modulus of Eq. (2.6) into classical transition amplitudes, where the paths $A_P^*$ and $A_Q$ are the same, and interference terms where the paths $A_P$ and $A_Q$ are distinct. Mathematically, this is expressed as:

$$P(A=j) = \sum_k |\sum_{P^{(j,n)}} A_P \langle k|\mathbb{B}_P^{(n)}|0\rangle|^2$$
$$= \sum_k \sum_{P^{(j,n)}} |A_P \langle k|\mathbb{B}_P^{(n)}|0\rangle|^2 + \sum_{P^{(j,n)} \neq Q^{(j,n)}} \sum_k \left( A_P^* A_Q \langle 0|\mathbb{B}_P|k\rangle\langle k|\mathbb{B}_Q^{(n)}|0\rangle \right). \tag{2.8}$$

The first term simplifies using the unitary matrix identity $\sum_k |\langle k|\mathbb{B}_P^{(n)}|0\rangle|^2 = 1$:

$$\sum_k \sum_{P^{(j,n)}} |A_P \langle k|\mathbb{B}_P^{(n)}|0\rangle|^2 = \sum_{P^{(j,n)}} |A_P|^2 \sum_k |\langle k|\mathbb{B}_P^{(n)}|0\rangle|^2 = \sum_{P^{(j,n)}} |A_P|^2. \tag{2.9}$$

Therefore the marginal probabilities for particle A become:

$$P(A=j) = \sum_{P^{(j,n)}} |A_P|^2 + \sum_{P^{(j,n)} \neq Q^{(j,n)}} \sum_k \left[ A_P^* A_Q \langle 0|\mathbb{B}_P^{(n)\dagger}|k\rangle\langle k|\mathbb{B}_Q^{(n)}|0\rangle \right]$$
$$= \sum_{P^{(j,n)}} |A_P|^2 + \sum_{P^{(j,n)} \neq Q^{(j,n)}} \left[ A_P^* A_Q \langle 0|\mathbb{B}_P^{(n)\dagger} \sum_k |k\rangle\langle k|\mathbb{B}_Q^{(n)}|0\rangle \right]$$
$$= \sum_{P^{(j,n)}} |A_P|^2 + \sum_{P^{(j,n)} \neq Q^{(j,n)}} \left[ A_P^* A_Q \langle 0|\mathbb{B}_P^{(n)\dagger} \mathbb{B}_Q^{(n)}|0\rangle \right], \tag{2.10}$$

where we have used the completeness relation $\sum_k |k\rangle\langle k| = \mathbb{I}$ to remove the dependence on the final states of particle B. We wish to express this marginal probability in terms of local propagation of information. To achieve this, $\langle 0|\mathbb{B}_P^{(n)\dagger}\mathbb{B}_Q^{(n)}|0\rangle$ in Eq. (2.10) can be sequentially decomposed such that it reflects the local acquisition of information by particle A's subsystem. Since $\mathbb{AB}_P^{(n)\dagger}\mathbb{AB}_Q^{(n)} = $ diag$\{e^{-i\theta_0^{(n)}(P)+i\theta_0^{(n)}(Q)}, e^{-i\theta_1^{(n)}(P)+i\theta_1^{(n)}(Q)}\} = $ diag$\{\bar{\eta}^{(n,0)}\eta^{(n,0)}, \bar{\eta}^{(n,1)}\eta^{(n,1)}\}$ we have:

$$\mathbb{AB}_P^{(n)\dagger}\mathbb{AB}_Q^{(n)} = \sum_k \left[ \mathbb{I} + \left( \bar{\eta}^{(n,k)}\eta^{(n,k)} - 1 \right) |k\rangle\langle k| \right], \tag{2.11}$$

Substituting this expansion into $\langle 0|\mathbb{B}_P^{(n)\dagger}\mathbb{B}_Q^{(n)}|0\rangle$ gives:

$$\begin{aligned}\langle 0|\mathbb{B}_P^{(n)\dagger}\mathbb{B}_Q^{(n)}|0\rangle =& \langle 0|\mathbb{B}_P^{(n-1)\dagger}\mathbb{B}^{(n)\dagger}\mathbb{A}\mathbb{B}_P^{(n)\dagger}\mathbb{A}\mathbb{B}_Q^{(n)}\mathbb{B}^{(n)}\mathbb{B}_Q^{(n-1)}|0\rangle \\ =& \langle 0|\mathbb{B}_P^{(n-1)\dagger}\mathbb{B}_Q^{(n-1)}|0\rangle \\ &+ \sum_k \left(\bar{\eta}^{(n,k)}\eta^{(n,k)} - 1\right)\langle 0|\mathbb{B}_P^{(n-1)\dagger}\mathbb{B}^{(n)\dagger}|k\rangle\langle k|\mathbb{B}^{(n)}\mathbb{B}_Q^{(n-1)}|0\rangle,\end{aligned}$$ (2.12)

where we have used $\mathbb{B}^{(n)\dagger}\mathbb{B}^{(n)} = \mathbb{I}$. The leading term $\langle 0|\mathbb{B}_P^{(n-1)\dagger}\mathbb{B}_Q^{(n-1)}|0\rangle$ in Eq. (2.12) is of the same form as the original term $\langle 0|\mathbb{B}_P^{(n)\dagger}\mathbb{B}_Q^{(n)}|0\rangle$ but for the transition amplitudes one layer earlier in the circuit. Therefore we can apply the same decomposition procedure to $\langle 0|\mathbb{B}_P^{(n-1)\dagger}\mathbb{B}_Q^{(n-1)}|0\rangle$ as was applied to $\langle 0|\mathbb{B}_P^{(n)\dagger}\mathbb{B}_Q^{(n)}|0\rangle$, which gives:

$$\begin{aligned}\langle 0|\mathbb{B}_P^{(n)\dagger}\mathbb{B}_Q^{(n)}|0\rangle =& \langle 0|\mathbb{B}_P^{(n-2)\dagger}\mathbb{B}_Q^{(n-2)}|0\rangle \\ &+ \sum_k \left(\bar{\eta}^{(n-1,k)}\eta^{(n-1,k)} - 1\right)\langle 0|\mathbb{B}_P^{(n-2)\dagger}\mathbb{B}^{(n-1)\dagger}|k\rangle\langle k|\mathbb{B}^{(n-1)}\mathbb{B}_Q^{(n-2)}|0\rangle \\ &+ \sum_k \left(\bar{\eta}^{(n,k)}\eta^{(n,k)} - 1\right)\langle 0|\mathbb{B}_P^{(n-1)\dagger}\mathbb{B}^{(n)\dagger}|k\rangle\langle k|\mathbb{B}^{(n)}\mathbb{B}_Q^{(n-1)}|0\rangle.\end{aligned}$$ (2.13)

The new leading term $\langle 0|\mathbb{B}_P^{(n-2)\dagger}\mathbb{B}_Q^{(n-2)}|0\rangle$ can itself be decomposed. The leading terms can be recursively decomposed in this way until $\langle 0|\mathbb{B}^{(1)\dagger}\mathbb{B}^{(1)}|0\rangle = 1$ is produced at the final stage of the decomposition. Performing this entire recursive decomposition gives:

$$\begin{aligned}\langle 0|\mathbb{B}_P^{(n)\dagger}\mathbb{B}_Q^{(n)}|0\rangle =& \langle 0|\mathbb{B}^{(1)\dagger}\mathbb{B}^{(1)}|0\rangle + \sum_t H_{P,Q}^{(t)} \\ =& 1 + \sum_{t=1}^n H_{P,Q}^{(t)},\end{aligned}$$ (2.14)

where $H_{P,Q}^{(t)} = \sum_k \left(\bar{\eta}^{(t,k)}\eta^{(t,k)} - 1\right)\langle 0|\mathbb{B}_P^{(t)\dagger}\mathbb{B}_Q^{(t)}|0\rangle$ are the terms which are split off during the decomposition of the leading terms at each layer $t$. Note that if no interaction occurs between particle A and B at layer $t$, then automatically $H_{P,Q}^{(t)}$ equals zero for this layer, because $\left(\bar{\eta}^{(t,k)}\eta^{(t,k)} - 1\right) = \left(e^{-i\theta_k^{(t)}(P)}e^{i\theta_k^{(t)}(Q)} - 1\right) = 0$ if $e^{-i\theta_k^{(t)}(P)}$ and $e^{i\theta_k^{(t)}(Q)}$ are both equal to one.

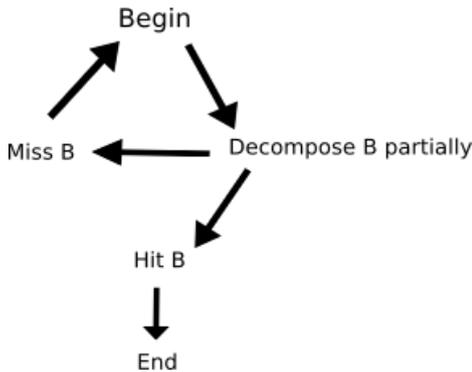

Figure 1: Flowchart of the decomposition procedure. At each partial decomposition, the initial term breaks into two pieces. The labels hit and miss refer to whether the corresponding piece contains the identity part (miss) or the interaction part (hit) of the partial decomposition.

The resulting marginal probability after the decomposition is:

$$P(A = j) = \sum_{P^{(j,n)}} |A_P|^2 + \sum_{P^{(j,n)} \neq Q^{(j,n)}} \left[ A_P^* A_Q \left( 1 + \sum_{t=1}^{n} H_{P,Q}^{(t)} \right) \right]$$
$$= \sum_{P^{(j,n)}} |A_P|^2 + \sum_{P^{(j,n)} \neq Q^{(j,n)}} \left[ A_P^* A_Q \lambda_{P,Q}^{(n)} \right], \tag{2.15}$$

where the hidden variables $\lambda_{P,Q}^{(n)}$ have been defined using $\lambda_{P,Q}^{(n)} \equiv 1 + \sum_{t=1}^{n} H_{P,Q}^{(n)}$. Eq. (2.15) can also be written in an alternative way:

$$P(A = j) = \sum_{P^{(j,n)}} |A_P|^2 + \sum_{P^{(j,n)} \neq Q^{(j,n)}} \left[ A_P^* A_Q \lambda_{P,Q}^{(n)} \right]$$
$$= \left( \sum_{P^{(j,n)}} |A_P|^2 + \sum_{P^{(j,n)} \neq Q^{(j,n)}} [A_P^* A_Q] \right) + \sum_{P^{(j)} \neq Q^{(j)}} \left[ A_P^* A_Q \left( \lambda_{P,Q}^{(n)} - 1 \right) \right]$$
$$= \sum_{P^{(j,n)}} |\mathbb{A}|^2 + \sum_{P^{(j,n)} \neq Q^{(j,n)}} \left[ A_P^* A_Q \left( \lambda_{P,Q}^{(n)} - 1 \right) \right], \tag{2.16}$$

where the matrix $\mathbb{A} = \mathbb{A}^{(n)} \mathbb{A}^{(n-1)} ... \mathbb{A}^{(1)}$ is the single-particle unitary matrix for particle A in the absence of interactions with particles external to the subsystem. This second form of the decomposition (Eq. (2.16)) is useful in isolating the non-zero contributions to the hidden variables $\lambda_{P,Q}^{(n)}$, which arise due to interactions.

## 2.3 Interpreting the decomposition

The physical picture described by this decomposition is summarized below:

1. The paths of particle A interact with particle B several times over their history.

2. Each time an interaction occurs, a hit of information $H_{P,Q}^{(t)}$ is obtained by the path of particle A undergoing the interaction. These hits of information are the products of two types transition amplitudes for particle B to arrive at the interaction location; amplitudes in which particle B is conditioned to interact with path $P$ of particle A, and amplitudes in which particle B is conditioned to interact with path $Q$ of particle A. The path of particle A which interacts with particle B is able to locally sample $H_{P,Q}^{(t)}$.

3. The hidden variables $\lambda_{P,Q}^{(t)}$ are recursively updated after each hit of information is acquired, according to the relation $\lambda_{P,Q}^{(t)} = \lambda_{P,Q}^{(t-1)} + H_{P,Q}^{(t)}$. Therefore it is not necessary for particle A to store the information $H_{P,Q}^{(t)}$ directly, but rather just the current value of the hidden variable $\lambda_{P,Q}^{(t)}$.

4. The final equation of the interpretation, Eq. (2.15), describes a sum-over-paths where the interference terms between paths are either enhanced or suppressed by the hidden variables $\lambda_{P,Q}^{(n)}$. For comparison, if $\lambda_{P,Q}^{(n)} = 1$ for all $P, Q$ then Eq. (2.15) reduces an ordinary sum over paths for particle A, where any interactions with the external system are ignored.

## 2.4 EPR-B experiment

Suppose that a Bell state is prepared by the application of a controlled not gate to the unentangled state $\frac{1}{\sqrt{2}} (|0\rangle_A + |1\rangle_A) |0\rangle_B$, where the kets $|j\rangle_A$ and $|k\rangle_B$ represent spatial modes of particle A and B

respectively. An EPR-B experiment can be performed upon this state by applying the local single-particle gates $\mathbb{A}^{(2)}$ and $\mathbb{B}^{(2)}\mathbb{H}$. These single particle gates essentially choose a measurement basis for particles A and B. We have included a Hadamard gate $\mathbb{H}$ along with the matrix $\mathbb{B}^{(2)}$ without loss of generality. Using the following decomposition of the CNOT gate in terms of a controlled phase gate, $C_{NOT}(A \to B) = (\mathbb{I} \otimes \mathbb{H})\left(\mathbb{A}\mathbb{B}^{(1)}\right)(\mathbb{I} \otimes \mathbb{H})$ where $\mathbb{A}\mathbb{B}^{(1)} = \text{diag}\{1,1,1,-1\}$, the final state of the system can then be expressed as:

$$\left(\mathbb{A}^{(2)} \otimes \mathbb{B}^{(2)}\right)\frac{1}{\sqrt{2}}\left(|0\rangle|0\rangle + |1\rangle|1\rangle\right) = \left(\mathbb{A}^{(2)} \otimes \mathbb{B}^{(2)}\mathbb{H}\right)C_{NOT}(A \to B)\frac{1}{\sqrt{2}}\left(|0\rangle + |1\rangle\right)|0\rangle$$

$$= \left(\mathbb{A}^{(2)} \otimes \mathbb{B}^{(2)}\mathbb{H}\right)C_{NOT}(A \to B)(\mathbb{H} \otimes \mathbb{I})|0,0\rangle$$

$$= \left(\mathbb{A}^{(2)} \otimes \mathbb{B}^{(2)}\mathbb{H}\right)(\mathbb{I} \otimes \mathbb{H})\left(\mathbb{A}\mathbb{B}^{(1)}\right)(\mathbb{I} \otimes \mathbb{H})(\mathbb{H} \otimes \mathbb{I})|0,0\rangle$$

$$= \left(\mathbb{A}^{(2)} \otimes \mathbb{B}^{(2)}\mathbb{H}\mathbb{H}\right)\left(\mathbb{A}\mathbb{B}^{(1)}\right)(\mathbb{H} \otimes \mathbb{H})|0,0\rangle \quad (2.17)$$

$$= \left(\mathbb{A}^{(2)} \otimes \mathbb{B}^{(2)}\right)\left(\mathbb{A}\mathbb{B}^{(1)}\right)(\mathbb{H} \otimes \mathbb{H})|0,0\rangle. \quad (2.18)$$

We have used $\mathbb{H}\mathbb{H} = \mathbb{I}$ in line (2.17) to remove these two Hadamard gates. The corresponding quantum circuit is shown in figure 2:

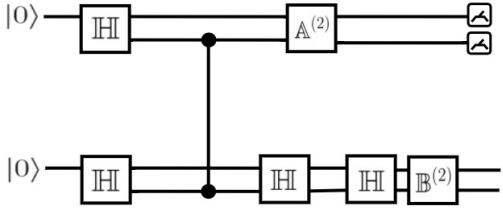

Figure 2: Circuit for the EPR-B experiment based on a spatially entangled Bell state.

There are only two paths of particle A which begin in mode $|0\rangle$ and end in mode $|j\rangle$, where $j \in \{0,1\}$. These paths have amplitudes $A^{(2)}_{j0}A^{(1)}_{00} = \frac{1}{\sqrt{2}}A^{(2)}_{j0}$ and $A^{(2)}_{j1}A^{(1)}_{10} = \frac{1}{\sqrt{2}}A^{(2)}_{j1}$ respectively, where we have taken $\mathbb{A}^{(1)} \equiv \mathbb{H}$ to give $A^{(1)}_{00} = A^{(1)}_{10} = \frac{1}{\sqrt{2}}$. Only the second path undergoes an interaction, so there is only one hit of information to account for. Eq. (2.16) tells us that the marginal probabilities are equal to:

$$P(A = j) = |\langle j|\mathbb{A}^{(2)}\mathbb{H}|0\rangle|^2 + 2\text{Re}\left\{\left(\frac{1}{\sqrt{2}}A^{(2)}_{j0}\right)^*\left(\frac{1}{\sqrt{2}}A^{(2)}_{j1}\right)\left(e^{i\theta_1} - 1\right)|B^{(1)}_{10}|^2\right\}, \quad (2.19)$$

In this particular case, $|B^{(1)}_{10}|^2 = \frac{1}{2}$ since we have defined $\mathbb{B}^{(1)} = \mathbb{H}$, and the controlled phase applied is $e^{i\theta_1} = -1$. Substituting these values into the marginal probability gives:

$$P(A = j) = |\langle j|\mathbb{A}^{(2)}\mathbb{H}|0\rangle|^2 - \text{Re}\left\{A^{(2)*}_{j0}A^{(2)}_{j1}\right\}$$

$$= |\langle j|\mathbb{A}^{(2)}\frac{1}{\sqrt{2}}\left(|0\rangle + |1\rangle\right)|^2 - \text{Re}\left\{A^{(2)*}_{j0}A^{(2)}_{j1}\right\}$$

$$= \frac{1}{2}|\langle j|\mathbb{A}^{(2)}|0\rangle|^2 + \frac{1}{2}|\langle j|\mathbb{A}^{(2)}|1\rangle|^2 + \frac{1}{2}\left(2\text{Re}\{\langle j|\mathbb{A}^{(2)}|0\rangle\langle j|\mathbb{A}^{(2)}|1\rangle\}\right) - \text{Re}\left\{A^{(2)*}_{j0}A^{(2)}_{j1}\right\}$$

$$= \frac{1}{2}|\langle j|\mathbb{A}^{(2)}|0\rangle|^2 + \frac{1}{2}|\langle j|\mathbb{A}^{(2)}|1\rangle|^2, \quad (2.20)$$

since $\text{Re}\{\langle j|\mathbb{A}^{(2)}|0\rangle\langle j|\mathbb{A}^{(2)}|1\rangle\} = \text{Re}\left\{A^{(2)*}_{j0}A^{(2)}_{j1}\right\}$. This is the known result for the marginal probabilities of an EPR-B experiment. In the local sum-over-paths interpretation, the interference terms

between the paths passing through modes $|0\rangle_A$ and $|1\rangle_A$ of Alice's circuit are completely turned off by the hidden variables. Note that these marginal probabilities are more commonly interpreted in a reduced density matrix ontology as arising from a classical sum of two density matrices for the subsystem, $|0\rangle_A\langle 0|_A$ and $|1\rangle_A\langle 1|_A$ respectively. However, the reduced density matrix ontology breaks down in more complicated circuits, while the local sum-over-paths remains valid.

## 2.5 Summary of results

In this section, we have made progress in addressing the concern that quantum subsystems are not described in a local way. It was shown that the marginal probabilities of these subsystems can be described in a sum-over-paths ontology, where local hidden variables modulate the amount of interference occurring between paths of the subsystem. The hidden variables update additively, which reduces the ontological complexity of the model. The essential feature of the model is that the paths of particle A undergoing the interaction extract information about particle B. This information is not grounded in the tensor product structure defining the interactions. Consequently, we have built a local ontology which accounts for the dissociation between information flow in quantum subsystems and the tensor product structure of the wavefunction, which was highlighted in paper one [1] of this series.

# 3 Beyond the two-particle case

The local sum-over-paths approach can be generalized to subsystems of N-particle systems. However, as the number of particles in the total system increases, the complexity of the ontology increases substantially. Therefore we will demonstrate the approach in the three-particle case and indicate how the method works for arbitrarily many particles.

## 3.1 The three-particle case

In the three particle case, the local sum-over-paths approach starts by representing the marginal probabilities in terms of a standard sum-over-paths:

$$P_A(j) = \sum_k \sum_l |\sum_{P^{(j,n)}} \sum_{M^{(k,n)}} \sum_{S^{(l,n)}} A_P B_M C_S e^{i\alpha(P,M)} e^{i\alpha(P,S)} e^{i\alpha(M,S)}|^2, \qquad (3.1)$$

where $A_P$, $B_M$ and $C_S$ are single particle path amplitudes for paths $P$, $M$ and $S$. As previously, we use indices on the paths of the summation $P^{(j,n)}$ to qualify properties about the paths which are to be summed. For instance in (3.1), the summations $\sum_{P^{(j,n)}} \sum_{M^{(k,n)}} \sum_{S^{(l,n)}}$ occur over paths for particle A, B and C which lead to modes $j$, $k$ and $l$ at layer $n$ of the circuit respectively. In this expression there are now three phase terms to contend with. The phase $e^{i\alpha(P,M)}$ is due to controlled phase gates between particles A and B, the phase $e^{i\alpha(P,S)}$ is due to controlled phase gates between particles A and C, and the phase $e^{i\alpha(M,S)}$ is due to controlled phase gates between particles B and C. To analyze the marginals further, we identify that $\sum_{M^{(k,n)}} \sum_{S^{(l,n)}} B_M C_S e^{i\alpha(P,M)} e^{i\alpha(P,S)} e^{i\alpha(M,S)} = \langle k,l|\mathbb{U}_P^{(n)}|0\rangle$, where $\mathbb{U}_P^{(n)}$ is a two particle unitary matrix giving the transition amplitudes of particles B and C up to layer $n$ of the circuit, given that particle A traverses path $P$. The initial state $|0\rangle$ is shorthand for $|0\rangle \equiv |0\rangle_A \otimes |0\rangle_B$. Having established these definitions, the marginal probabilities can be expressed as:

$$P(A=j) = \sum_k \sum_l |\sum_{P^{(j,n)}} A_P \left( \sum_{M^{(k,n)}} \sum_{S^{(l,n)}} B_M C_S e^{i\alpha(P,M)} e^{i\alpha(P,S)} e^{i\alpha(M,S)} \right)|^2$$

$$= \sum_k \sum_l |\sum_{P^{(j,n)}} A_P \langle k,l| \mathbb{U}_P^{(n)} |0\rangle|^2$$

$$= \sum_k \sum_l \sum_{P^{(j,n)}} |A_P \langle k,l| \mathbb{U}_P^{(n)} |0\rangle|^2 + \sum_k \sum_l \sum_{P^{(j,n)} \neq Q^{(j,n)}} \left[ A_P^* A_Q \langle 0| \mathbb{U}_P^{(n)\dagger} |k,l\rangle \langle k,l| \mathbb{U}_Q^{(n)} |0\rangle \right]$$

$$= \sum_{P^{(j,n)}} |A_P|^2 \sum_k \sum_l |\langle k,l| \mathbb{U}_P^{(n)} |0\rangle|^2 + \sum_{P^{(j,n)} \neq Q^{(j,n)}} \left[ A_P^* A_Q \langle 0| \mathbb{U}_P^{(n)\dagger} \sum_k \sum_l |k,l\rangle \langle k,l| \mathbb{U}_Q^{(n)} |0\rangle \right]$$

$$= \sum_{P^{(j,n)}} |A_P|^2 + \sum_{P^{(j,n)} \neq Q^{(j,n)}} \left[ A_P^* A_Q \langle 0| \mathbb{U}_P^{(n)\dagger} \mathbb{U}_Q^{(n)} |0\rangle \right], \tag{3.2}$$

where we have used the unitary matrix identity $\sum_k \sum_l |\langle k,l| \mathbb{U}_P^{(n)} |0,0\rangle|^2 = 1$ and the completeness relation $\sum_k \sum_l |k,l\rangle\langle k,l| = \sum_k |k\rangle\langle k| \otimes \sum_l |l\rangle\langle l| = \mathbb{I} \otimes \mathbb{I}$ to remove the dependence on the final states of particles B and C. Just as in the two particle case, the goal is to decompose $\langle 0| \mathbb{U}_P^{(n)\dagger} \mathbb{U}_Q^{(n)} |0\rangle$ such that it reflects the local propagation of information into particle A's subsystem. The decomposition procedure for $\langle 0| \mathbb{U}_P^{(n)\dagger} \mathbb{U}_Q^{(n)} |0\rangle$ is provided in Appendix A. The result of the decomposition of $\langle 0| \mathbb{U}_P^{(n)\dagger} \mathbb{U}_Q^{(n)} |0\rangle$ is the following:

$$\langle 0| \mathbb{U}_P^{(n)\dagger} \mathbb{U}_Q^{(n)} |0\rangle = 1 + \sum_{r=1}^n H_{P,Q}^{(r)} \equiv \lambda_{P,Q}^{(n)}, \tag{3.3}$$

where $\lambda_{P,Q}^{(n)}$ are hidden variables for paths $P$ and $Q$ of particle A. As shown in Appendix A, the hits of information $H_{P,Q}^{(r)}$ are defined by:

$$H_{P,Q}^{(r)} \equiv \sum_k \sum_{M^{(k,t)}} \sum_{N^{(k,t)}} \left[ \delta_{P,Q,M,N}^{(r)} \left( 1 + \sum_{t=1}^{r-1} \gamma_{P,Q,M,N}^{(t)} + \sum_{t=1}^r \chi_{P,Q,M,N}^{(t)} \right) \right] \tag{3.4}$$

$$+ \sum_l \sum_{S^{(l,t)}} \sum_{T^{(l,t)}} \left[ \delta_{P,Q,S,T}^{(r)} \left( 1 + \sum_{t=1}^r \gamma_{P,Q,S,T}^{(t)} + \sum_{t=1}^r \chi_{P,Q,S,T}^{(t)} \right) \right],$$

where $\delta_{P,Q,M,N}^{(r)}$, $\delta_{P,Q,S,T}^{(r)}$, $\gamma_{P,Q,M,N}^{(t)}$, $\gamma_{P,Q,S,T}^{(t)}$, $\chi_{P,Q,M,N}^{(t)}$ and $\chi_{P,Q,S,T}^{(t)}$ are all distinct types of hidden variables which will be explained in more detail. The main point is that the marginal probabilities are again given by a sum over paths, where the interference terms between paths are modulated by the hidden variables. Explicitly, we can write the marginal probabilities as:

$$P(A=j) = \sum_{P^{(j,n)}} |A_P|^2 + \sum_{P^{(j,n)} \neq Q^{(j,n)}} \left[ A_P^* A_Q \left( \lambda_{P,Q}^{(n)} \right) \right], \tag{3.5}$$

where $\lambda_{P,Q}^{(n)} \equiv 1 + \sum_{r=1}^n H_{P,Q}^{(r)}$.

## 3.2 Interpreting the decomposition

The resulting sum-over-paths interpretation can be understood in the following way:

1. The paths of particle A repeatedly interact with the paths of particles B and C.

2. Each time an interaction occurs, a hit of information $H_{P,Q}^{(t)}$ is obtained by the path of particle A undergoing the interaction. These hits of information reflect the possible ways in which information can flow into particle A's subsystem.

   (a) Suppose that an A-B interaction occurs at layer r of the circuit. Prior to the A-B interaction, particle A can interact with particle C, and recursively store the hidden variables $\sum_{t=1}^{r-1} \gamma_{P,Q,M,N}^{(t)}$. In addition, prior to the A-B interaction, particle B can interact with particle C and recursively store the hidden variables $\sum_{t=1}^{r} \chi_{P,Q,M,N}^{(t)}$. All of this information is then weighted by the factor $\delta_{P,Q,M,N}^{(r)}$ which is obtained during the A-B interaction to form the hit of information $H_{P,Q}^{(r)}$. Particle A does this for all paths $M, N$ of particle B which participate in the interaction.

   (b) Suppose instead that an A-C interaction occurs at layer r of the circuit. Prior to the A-C interaction, particle A can interact with particle B and recursively store the hidden variables $\sum_{t=1}^{r} \gamma_{P,Q,S,T}^{(t)}$. Furthermore, prior to the A-C interaction, particle C can interact with particle B and recursively store the hidden variables $\sum_{t=1}^{r} \chi_{P,Q,S,T}^{(t)}$. All of this information is then weighted by $\delta_{P,Q,S,T}^{(r)}$ which is obtained during the A-C interaction to form the hit of information $H_{P,Q}^{(r)}$. Particle A does this for all paths $S$ and $T$ of particle C which participate in the interaction.

   (c) Of course, both an A-B interaction and an A-C interaction can occur at layer r, in which case $\lambda_{P,Q}^{(t)}$ is updated twice at this layer. However, we have defined the order of gates (without loss of generality) that the A-C interaction at layer r always occurs after the A-B interaction at layer r. This is why $\sum_{t=1}^{r-1} \gamma_{P,Q,M,N}^{(t)}$ is summed to $r-1$ while $\sum_{t=1}^{r} \gamma_{P,Q,S,T}^{(t)}$ is summed to r.

3. The hidden variables $\lambda_{P,Q}^{(t)}$ are recursively updated after each hit of information is acquired, according to the relation $\lambda_{P,Q}^{(t)} = \lambda_{P,Q}^{(t-1)} + H_{P,Q}^{(t)}$.

4. The hidden variables $\lambda_{P,Q}^{(n)}$ modulate the interference occurring between the path amplitudes of paths $P$ and $Q$ of particle A.

## 3.3 The N-particle case

The method of analysis developed for the three-particle case can be generalized to more complex situations. Suppose that we have an N-particle system, and wish to describe an M-particle subset of the total system as a quantum subsystem. A simple way to extend the local sum-over-paths approach to this situation is to condition on paths of the subsystem which are now configuration space paths, rather than single-particle paths as was done previously.

Consider the configuration space path $P$, which is formed by single particle paths $P_1, P_2, ..., P_M$ of an M-particle subsystem. It has path amplitudes of the form $A_P = A_{P_1} A_{P_2} ... A_{P_M} e^{i\alpha(P_1,...,P_M)}$ where $A_{P_i}$ denote single particle path amplitudes and $e^{i\alpha(P_1,...,P_M)}$ is the phase acquired due to any controlled phase gates between particle of the subsystem. The marginal probabilities for the subsystem are then given by:

$$P(j_1, j_2, ..., j_M) = \sum_{k_1} ... \sum_{k_{N-M}} | \sum_{P(j_1,...,j_M,n)} A_P \langle k_1,...,k_{N-M} | \mathbb{U}_P | 0 \rangle |^2, \tag{3.6}$$

where $j_1, ..., j_M$ label spatial modes of particles of the subsystem and $k_1, ..., k_{N-M}$ label spatial modes of particles of the external system. The unitary matrix $\mathbb{U}_P$ describes the time evolution of the external system, given that the particles of the subsystem collectively traverse configuration space path $P$. Using a local sum-over-paths decomposition, this expression can be reduced to:

$$\begin{aligned} P(j_1, j_2, ..., j_M) &= \sum_{k_1} ... \sum_{k_{N-M}} |\sum_{P(j_1,,,j_M,n)} A_P \langle k_1, ..., k_{N-M} | \mathbb{U}_P | 0 \rangle|^2 \\ &= \sum_{P(j_1,,,j_M,n)} |A_P|^2 + \sum_{P(j_1,,,j_M,n) \neq Q(j_1,,,j_M,n)} 2Re\{A_P^* A_Q \langle 0 | \mathbb{U}_P^\dagger \mathbb{U}_Q | 0 \rangle\} \\ &= \sum_{P(j_1,,,j_M,n)} |A_P|^2 + \sum_{P(j_1,,,j_M,n) \neq Q(j_1,,,j_M,n)} 2Re\{A_P^* A_Q \lambda_{P,Q}^{(n)}\}, \end{aligned} \quad (3.7)$$

where $\langle 0 | \mathbb{U}_P^\dagger \mathbb{U}_Q | 0 \rangle = \lambda_{P,Q}^{(n)}$ has been decomposed into hidden variables $\lambda_{P,Q}^{(n)} = 1 + \sum_{r=1}^{n} H_{P,Q}^{(r)}$. When the external system has a large number of particles, the expressions for hits of information $H_{P,Q}^{(r)}$ will be quite complex. This is due to a large number of ways in which information can propagate into the subsystem. Information can be shared extensively between paths of the external system before reaching the subsystem. Nevertheless, it is evident that $\langle 0 | \mathbb{U}_P^\dagger \mathbb{U}_Q | 0 \rangle$ can still be decomposed in a local way. This is done by conditioning upon the paths of all particles but one, then expressing the remaining paths of this particle in matrix notation to perform a partial decomposition. This is done until all particles have been hit by the interaction parts of the partial decompositions.

If we take M=1, Eq. (3.7) represents the local sum-over-paths interpretation for a single-particle subsystem with N-1 particles in the external system. Of course, when M>1 the resulting local sum-over-paths interpretation is only somewhat local. It describes the local acquisition of information from the external system, but still requires non-locality to explain the correlations between particles of the subsystem itself. However, this is still an achievement beyond the standard quantum formalism, which provides an entirely non-local description of the subsystem.

## 3.4 Applications

### 3.4.1 Event horizons

There are areas in which a (somewhat) local description of multipartite subsystems can play an important explanatory role. For instance when particles of the external system cross event horizons, the local sum-over-paths approach can be used explain the measurement results of particles of a subsystem, taken to comprise of particles on the observable side of the horizon. A traditional quantum interpretation requires non-local effects to propagate from behind the event horizon.

To provide an example, suppose that we have a collection of entangled particles which have interacted in a quantum circuit. Some of these particles are not measured, but are released into the black hole. We can define the particles still on the observable side of the black hole as being part of a multipartite subsystem. The local sum-over-paths interpretation then describes the multipartite subsystem in a local way. In a more extreme example, we can define the quantum subsystem to consist of all of the particles in the universe which are not inside the black hole.

There are several reasons why the local ontology is preferable. First, the mechanism for non-local influences is unknown, and it is open to speculation whether these influences could survive in the strong gravitational regime. A local ontology of subsystems avoids these difficult questions. Second, there is ongoing debate regarding whether information is destroyed within black holes. The local sum-over-paths interpretation provides a local, non-unitary description of the subsystem which can account for the destruction of information in the external system.

Another use for the local sum-over-paths approach arises in cosmology, where entangled particles can be distributed either side of a cosmological event horizon. As in the case of the black hole event horizon, the interpretation can provide a local description in specific thought experiments, or alternatively can be used as part of the ontology for the entire world on the observable side of the horizon. In the case of cosmological horizons, the local ontology avoids the question of whether the purported non-local influences affecting subsystems are immune to the expansion of space.

### 3.4.2 Ontological complexity and new physics

The local interpretation of N-particle systems suggests that particles store an excessively large amount of information about their interactions. The ontological complexity of this model is quite peculiar, particularly in situations where decoherence has occurred and the net result of this information is just to wash out quantum interference effects.

Perhaps there are ways in which quantum theory can be modified such that the ontological complexity of the model is reduced. In this respect, the local sum-over-paths interpretation is a good base theory from which to search for such modifications. One suggestion is to test whether there are limits on the storage or propagation of these hidden variables. For instance, there may be limits with respect to storage duration or capacity of these variables. Or alternatively, there may be limits with respect number of times these hidden variables can be shared between different particles. Such limits can modify the quantum predictions in useful ways which have not been tested experimentally.

## 3.5 Summary of results

This section has described how the local sum-over-paths approach can be extended to the three-particle and N-particle cases. Beyond the two-particle case, deriving a local interpretation of quantum subsystems is quite challenging. This is because of the multitude of ways in which information can enter the subsystem. To describe the transmission of this information in a local manner, we have postulated a network of local hidden variables and update rules. These information contained in these hidden variables is shared during interactions.

A local description of subsystems in the N-particle case can prove quite useful. We have highlighted two areas of application. First, the local interpretation can provide explanations for quantum predictions in situations involving event horizons. Second, the local interpretation can be used as a base from which to search for deviations from the quantum predictions. Paper [2] will describe how the local sum-over-paths interpretation can be used in more comprehensive ontologies of quantum mechanics.

# 4 Conclusion

The world of quantum subsystems is a strange and complex one. While the wavefunction for the total system has a fairly straightforward[1] Schrödinger time evolution equation, to restore locality to quantum subsystems, and intricate network of local hidden variables is required. This paper has captured the essence of how these hidden variables are stored, propagated and updated to produce the marginal probabilities of the subsystem. We have demonstrated the local sum-over-paths ontology in the two and three-particle cases.

The essence of the interpretation is that paths of the subsystem obtain information about the external system during interactions. This information is extracted in a manner which is discordant with the tensor product structure of the wavefunction. Consequently, it is necessary to go beyond the paradigm of Schrödinger time evolution to describe how this information is propagated into the subsystem.

The local sum-over-paths interpretation not only resolves the issue of locality of quantum subsystems, it resolves some very fundamental questions in standard quantum mechanics where subsystems are the only quantum systems which can be observed, such as in the presence of event horizons. The interpretation furthermore provides a new perspective quantum information and decoherence.

---

[1] Albeit in configuration space.

# A  Decomposition procedure in the three particle case

In this section, we will describe how to decompose $\langle 0|\mathbb{U}_P^{(n)\dagger}\mathbb{U}_Q^{(n)}|0\rangle$ such that it expresses a local flow of information in the three-particle case. The following chart shows the decomposition procedure in graphical form.

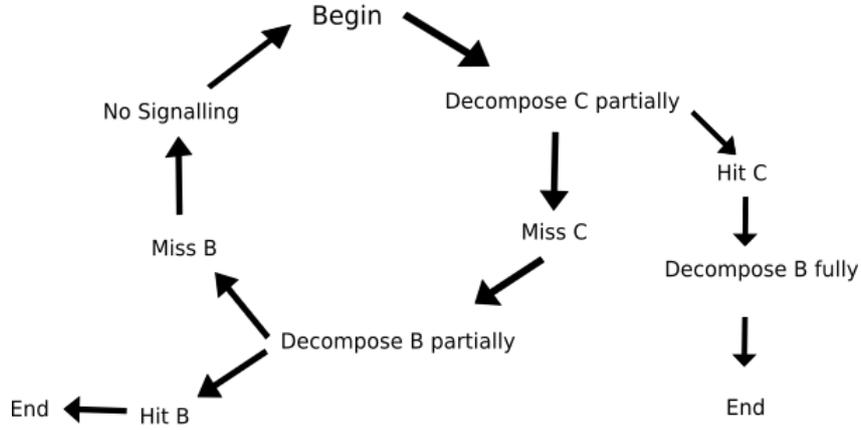

Figure 3: Recursive decomposition of $\langle 0|\mathbb{U}_P^{(n)\dagger}\mathbb{U}_Q^{(n)}|0\rangle$. At each partial decomposition, the initial term breaks into two pieces. The labels hit and miss refer to whether the corresponding piece contains the identity part (miss) or the interaction part (hit) of the partial decomposition. If the cycle begins at layer $t$, after the no-signaling condition is used to remove any irrelevant B-C interactions, the cycle will begin again at layer $t-1$.

## A.1  Definitions

Expressed as a sum over paths, the term $\langle 0|\mathbb{U}_P^{(n)\dagger}\mathbb{U}_Q^{(n)}|0\rangle$ is:

$$\langle 0|\mathbb{U}_P^{(n)\dagger}\mathbb{U}_Q^{(n)}|0\rangle = \sum_{P^{(j,n)} \neq Q^{(j,n)}} B_M^* B_N C_S^* C_T e^{-i\alpha(P,M)} e^{i\alpha(Q,N)} e^{-i\alpha(P,S)} e^{i\alpha(Q,T)} e^{-i\alpha(M,S)} e^{i\alpha(N,T)}$$

$$= \sum_{P^{(j,n)} \neq Q^{(j,n)}} B_M^* B_N C_S^* C_T \, \underrightarrow{\bar{\eta}}^{(n)} \, \underrightarrow{\eta}^{(n)} \, \underrightarrow{\bar{\tau}}^{(n)} \, \underrightarrow{\tau}^{(n)} \, \underrightarrow{\bar{\varepsilon}}^{(n)} \, \underrightarrow{\varepsilon}^{(n)} \qquad (A.1)$$

For brevity we have written the following phases in a condensed notation:

$$\underrightarrow{\bar{\eta}}^{(n)} \equiv e^{-i\alpha(P,M)} \equiv \bar{\eta}^{(n)} \, \underrightarrow{\bar{\eta}}^{(n-1)} \qquad \underrightarrow{\eta}^{(n)} \equiv e^{i\theta(Q,N)} \equiv \eta^{(n)} \, \underrightarrow{\eta}^{(n-1)}$$

$$\underrightarrow{\bar{\tau}}^{(n)} \equiv e^{-i\theta(P,S)} \equiv \bar{\tau}^{(n)} \, \underrightarrow{\bar{\tau}}^{(n-1)} \qquad \underrightarrow{\tau}^{(n)} \equiv e^{i\alpha(Q,T)} \equiv \tau^{(n)} \, \underrightarrow{\tau}^{(n-1)}$$

$$\underrightarrow{\bar{\varepsilon}}^{(n)} \equiv e^{-i\alpha(M,S)} \equiv \bar{\varepsilon}^{(n)} \, \underrightarrow{\bar{\varepsilon}}^{(n-1)} \qquad \underrightarrow{\varepsilon}^{(n)} \equiv e^{i\alpha(N,T)} \equiv \varepsilon^{(n)} \, \underrightarrow{\varepsilon}^{(n-1)}$$

The phases are defined recursively using $\underrightarrow{\eta}^{(n)} = \eta^{(n)}\underrightarrow{\eta}^{(n-1)}$ for instance. The symbol $\eta^{(n)}$ denotes the phase due to A-B interactions occurring at layer $n$ of the circuit. The symbol $\underrightarrow{\eta}^{(n-1)}$ is the cumulative product of all of the phases due to A-B interactions occurring up to layer $n-1$ of the circuit. These symbols implicitly have a dependence on the configuration space paths chosen.

## A.2 Begin recursive sequence → Decompose C partially

Start at layer n of the circuit. Transform $\langle 0|\mathbb{U}_P^{(n)\dagger}\mathbb{U}_Q^{(n)}|0\rangle$ into matrix amplitudes for particle C given that particles A and B take paths $P$ and $M$ (or $Q$ and $N$) respectively. Then collect these matrix amplitudes together. Explicitly, we have:

$$\langle 0|\mathbb{U}_P^{(n)\dagger}\mathbb{U}_Q^{(n)}|0\rangle = \sum_{k,l}\langle 0|\mathbb{U}_P^{(n)\dagger}|k,l\rangle\langle k,l|\mathbb{U}_Q^{(n)}|0\rangle$$

$$=\sum_k\sum_l\sum_{M^{(k,n)}}\left(B_M^*\,\underrightarrow{\bar\eta}^{(n)}\langle 0|\mathbb{C}_{P,M}^{(n)\dagger}|l\rangle\right)\sum_{N^{(k,n)}}\left(B_N\,\underrightarrow{\eta}^{(n)}\langle l|\mathbb{C}_{Q,N}^{(n)}|0\rangle\right)$$

$$=\sum_k\sum_{M^{(k,n)}}\sum_{N^{(k,n)}}B_M^*B_N\,\underrightarrow{\bar\eta}^{(n)}\,\underrightarrow{\eta}^{(n)}\langle 0|\mathbb{C}_{P,M}^{(n)\dagger}\left(\sum_l|l\rangle\langle l|\right)\mathbb{C}_{Q,N}^{(n)}|0\rangle \quad (A.2)$$

$$=\sum_k\sum_{M^{(k,n)}}\sum_{N^{(k,n)}}B_M^*B_N\,\underrightarrow{\bar\eta}^{(n)}\,\underrightarrow{\eta}^{(n)}\langle 0|\mathbb{C}_{P,M}^{(n)\dagger}\mathbb{C}_{Q,N}^{(n)}|0\rangle, \quad (A.3)$$

where we have used the completeness relation $\sum_l|l\rangle\langle l| = 1$ in line (A.2). The matrix $\mathbb{C}_{P,M}^{(t)} \equiv \mathbb{AC}_P^{(t)}\mathbb{BC}_M^{(t)}\mathbb{C}^{(t)}...\mathbb{AC}_P^{(1)}\mathbb{BC}_M^{(1)}\mathbb{C}^{(1)}$ denotes the unitary matrix for particle C up to layer t, given particles A and B take paths $P$ and $M$. We can express $\mathbb{C}_{P,M}^{(t)}$ recursively as $\mathbb{C}_{P,M}^{(t)} = \mathbb{AC}_P^{(t)}\mathbb{BC}_M^{(t)}\mathbb{C}^{(t)}\mathbb{C}_{P,M}^{(t-1)}$. The matrices $\mathbb{AC}_P^{(t)}$ describe the effect of the controlled phase gate $\mathbb{AC}^{(t)}$ on particle C given that particle A takes path $P$, while the matrices $\mathbb{BC}_M^{(t)}$ describe the effect of the controlled phase gate $\mathbb{BC}^{(t)}$ on particle C given that particle B takes path M.

The the factor $\langle 0|\mathbb{C}_{P,N}^{(n)\dagger}\mathbb{C}_{Q,M}^{(n)}|0\rangle$ in Eq. (A.3) is decomposed a single time by inserting the following expansion:

$$\mathbb{AC}_P^{(n)\dagger}\mathbb{AC}_Q^{(n)} = \sum_l\left[\mathbb{I}+\left(\bar\tau^{(n)}\tau^{(n)}-1\right)|l\rangle\langle l|\right], \quad (A.4)$$

which gives:

$$\langle 0|\mathbb{C}_{P,M}^{(n)\dagger}\mathbb{C}_{Q,N}^{(n)}|0\rangle = \sum_{M^{(k,n)}}\sum_{N^{(k,n)}}\langle 0|\mathbb{C}_{P,M}^{(n-1)\dagger}\mathbb{C}^{(n)\dagger}\mathbb{BC}_M^{(n)\dagger}\mathbb{AC}_P^{(n)\dagger}\mathbb{AC}_Q^{(n)}\mathbb{BC}_N^{(n)}\mathbb{C}^{(n)}\mathbb{C}_{Q,N}^{(n-1)}|0\rangle$$

$$=\langle 0|\mathbb{C}_{P,M}^{(n-1)\dagger}\mathbb{C}^{(n)\dagger}\mathbb{BC}_M^{(n)\dagger}\mathbb{BC}_N^{(n)}\mathbb{C}^{(n)}\mathbb{C}_{Q,N}^{(n-1)}|0\rangle \quad (A.5)$$

$$+\sum_l\left(\bar\tau^{(n)}\tau^{(n)}-1\right)\langle 0|\mathbb{C}_{P,M}^{(n-1)\dagger}\mathbb{C}^{(n)\dagger}\mathbb{BC}_M^{(n)\dagger}|l\rangle\langle l|\mathbb{BC}_N^{(n)}\mathbb{C}^{(n)}\mathbb{C}_{Q,N}^{(n-1)}|0\rangle.$$

Therefore the inner product $\langle 0|\mathbb{U}_P^{(n)\dagger}\mathbb{U}_Q^{(n)}|0\rangle$ becomes:

$$\langle 0|\mathbb{U}_P^{(n)\dagger}\mathbb{U}_Q^{(n)}|0\rangle$$

$$=\sum_k\sum_{M^{(k,n)}}\sum_{N^{(k,n)}}B_M^*B_M^*\,\underrightarrow{\bar\eta}^{(n)}\,\underrightarrow{\eta}^{(n)}\langle 0|\mathbb{C}_{P,M}^{(n-1)\dagger}\mathbb{C}^{(n)\dagger}\mathbb{BC}_M^{(n)\dagger}\mathbb{BC}_N^{(n)}\mathbb{C}^{(n)}\mathbb{C}_{Q,N}^{(n-1)}|0\rangle \quad (A.6)$$

$$+\sum_l\sum_k\sum_{M^{(k,n)}}\sum_{N^{(k,n)}}\left(\bar\tau^{(n)}\tau^{(n)}-1\right)B_M^*B_M^*\,\underrightarrow{\bar\eta}^{(n)}\,\underrightarrow{\eta}^{(n)}\langle 0|\mathbb{C}_{P,M}^{(n-1)\dagger}\mathbb{C}^{(n)\dagger}\mathbb{BC}_M^{(n)\dagger}|l\rangle\langle l|\mathbb{BC}_N^{(n)}\mathbb{C}^{(n)}\mathbb{C}_{Q,N}^{(n-1)}|0\rangle.$$

There are two terms. The first corresponds to particle C being missed by the $|l\rangle\langle l|$ part of the expansion. The second corresponds to particle C being hit by the $|l\rangle\langle l|$ part of the expansion.

## A.3 Hit C → Decompose particle B fully → End

Now concentrate on decomposing the second term of Eq. (A.6). To proceed, express this term as a sum over paths, then collect the paths of particle B together to form a matrix representation:

$$\sum_k \sum_l \left( \bar{\tau}^{(n)} \tau^{(n)} - 1 \right) B_M^* B_M^* \bar{\underset{\rightarrow}{\eta}}^{(n)} \underset{\rightarrow}{\eta}^{(n)} \langle 0 | \mathbb{C}_{P,M}^{(n-1)\dagger} \mathbb{C}^{(n)\dagger} \mathbb{B} \mathbb{C}_M^{(n)\dagger} | l \rangle \langle l | \mathbb{B} \mathbb{C}_N^{(n)} \mathbb{C}^{(n)} \mathbb{C}_{Q,N}^{(n-1)} | 0 \rangle$$

$$= \sum_k \sum_l \left( \bar{\tau}^{(n)} \tau^{(n)} - 1 \right) B_M^* B_M^* \bar{\underset{\rightarrow}{\eta}}^{(n)} \underset{\rightarrow}{\eta}^{(n)} \left( \sum_{S^{(l,n)}} C_S^{(k)\dagger} \bar{\underset{\rightarrow}{\tau}}^{(n-1)} \bar{\underset{\rightarrow}{\varepsilon}}^{(n)} \right) \left( \sum_{T^{(l,n)}} C_T^{(k)} \underset{\rightarrow}{\tau}^{(n-1)} \underset{\rightarrow}{\varepsilon}^{(n)} \right)$$

$$= \sum_l \sum_{S^{(l,n)}} \sum_{T^{(l,n)}} \left( \bar{\tau}^{(n)} \tau^{(n)} - 1 \right) C_S^\dagger C_T \bar{\underset{\rightarrow}{\tau}}^{(n-1)} \underset{\rightarrow}{\tau}^{(n-1)} \sum_k \left( \sum_{M^{(k,n)}} B_M^* \bar{\underset{\rightarrow}{\eta}}^{(n)} \bar{\underset{\rightarrow}{\varepsilon}}^{(n)} \right) \left( \sum_{N^{(k,n)}} B_M^* \underset{\rightarrow}{\eta}^{(n)} \underset{\rightarrow}{\varepsilon}^{(n)} \right)$$

$$= \sum_l \sum_{S^{(l,n)}} \sum_{T^{(l,n)}} \left( \bar{\tau}^{(n)} \tau^{(n)} - 1 \right) C_S^\dagger C_T \bar{\underset{\rightarrow}{\tau}}^{(n-1)} \underset{\rightarrow}{\tau}^{(n-1)} \sum_k \langle 0 | \mathbb{B}_{P,S}^\dagger | k \rangle \langle k | \mathbb{B}_{Q,T} | 0 \rangle$$

$$= \sum_l \sum_{S^{(l,n)}} \sum_{T^{(l,n)}} \left( \bar{\tau}^{(n)} \tau^{(n)} - 1 \right) C_S^\dagger C_T \bar{\underset{\rightarrow}{\tau}}^{(n-1)} \underset{\rightarrow}{\tau}^{(n-1)} \langle 0 | \mathbb{B}_{P,S}^\dagger \mathbb{B}_{Q,T} | 0 \rangle. \tag{A.7}$$

Now decompose $\langle 0 | \mathbb{B}_{P,S}^\dagger \mathbb{B}_{Q,T} | 0 \rangle$ fully. Firstly note that:

$$\langle 0 | \mathbb{B}_{P,S}^\dagger \mathbb{B}_{Q,T} | 0 \rangle = \langle 0 | \mathbb{B}_{P,S}^{(n-1)\dagger} \mathbb{B}^{(n)\dagger} \mathbb{B} \mathbb{C}_S^{(n)\dagger} \mathbb{A} \mathbb{B}_P^{(n)\dagger} \mathbb{A} \mathbb{B}_Q^{(n)} \mathbb{B} \mathbb{C}_T^{(n)} \mathbb{B}^{(n)} \mathbb{B}_{Q,T}^{(n-1)} | 0 \rangle, \tag{A.8}$$

where $\mathbb{B}_{P,S}^{(t)}$ is the unitary matrix for particle B up to layer t, given that particles A and C traverse paths $P$ and $S$ respectively. The matrix $\mathbb{AB}_P^{(t)}$ describes the effect of controlled phase gate $\mathbb{AB}^{(t)}$ on particle B, given that particle A traverses path $P$. The matrix $\mathbb{BC}_T^{(n)}$ denotes the effect of controlled phase gate $\mathbb{BC}^{(n)}$ on particle B given that particle C traverses path T. The decomposition works by performing the following expansions:

$$\mathbb{AB}_P^{(n)\dagger} \mathbb{AB}_Q^{(n)} = \mathbb{I} + \sum_k \left( \bar{\eta}^{(n)} \eta^{(n)} - 1 \right) | k \rangle \langle k |. \tag{A.9}$$

$$\mathbb{BC}_S^{(n)\dagger} \mathbb{BC}_T^{(n)} = \mathbb{I} + \sum_k \left( \bar{\varepsilon}^{(n)} \varepsilon^{(n)} - 1 \right) | k \rangle \langle k |. \tag{A.10}$$

until either $\left( \bar{\eta}^{(t)} \eta^{(t)} - 1 \right) | k \rangle \langle k |$ or $\sum_k \left( \bar{\varepsilon}^{(t)} \varepsilon^{(t)} - 1 \right) | k \rangle \langle k |$ are hit at layer $t$. The result is that:

$$\langle 0 | \mathbb{B}_{P,S}^\dagger \mathbb{B}_{Q,T} | 0 \rangle = 1 + \sum_{t=1}^n \sum_k \left( \bar{\eta}^{(t)} \eta^{(t)} - 1 \right) \langle 0 | \mathbb{B}_{P,S}^{(t-1)\dagger} \mathbb{B}^{(t)\dagger} \mathbb{BC}_S^{(t)\dagger} | k \rangle \langle k | \mathbb{BC}_T^{(t)} \mathbb{B}^{(t)} \mathbb{B}_{Q,T}^{(t-1)} | 0 \rangle$$

$$+ \sum_{t=1}^n \sum_k \left( \bar{\varepsilon}^{(t)} \varepsilon^{(t)} - 1 \right) \langle 0 | \mathbb{B}_{P,S}^{(t-1)\dagger} \mathbb{B}^{(t)\dagger} | k \rangle \langle k | \mathbb{B}^{(t)} \mathbb{B}_{Q,T}^{(t-1)} | 0 \rangle$$

$$= 1 + \sum_{t=1}^n \gamma_{P,Q,T,S}^{(t)} + \sum_{t=1}^n \chi_{P,Q,S,T}^{(t)}, \tag{A.11}$$

where the hidden variables $\gamma_{P,Q,T,S}^{(t)}$ and $\chi_{P,Q,S,T}^{(t)}$ are defined as:

$$\gamma_{P,Q,T,S}^{(t)} \equiv \sum_k \left( \bar{\eta}^{(t)} \eta^{(t)} - 1 \right) \langle 0 | \mathbb{B}_{P,S}^{(t-1)\dagger} \mathbb{B}^{(t)\dagger} \mathbb{BC}_S^{(t)\dagger} | k \rangle \langle k | \mathbb{BC}_T^{(t)} \mathbb{B}^{(t)} \mathbb{B}_{Q,T}^{(t-1)} | 0 \rangle. \tag{A.12}$$

$$\chi_{P,Q,S,T}^{(t)} \equiv \sum_k \left( \bar{\varepsilon}^{(t)} \varepsilon^{(t)} - 1 \right) \langle 0 | \mathbb{B}_{P,S}^{(t-1)\dagger} \mathbb{B}^{(t)\dagger} | k \rangle \langle k | \mathbb{B}^{(t)} \mathbb{B}_{Q,T}^{(t-1)} | 0 \rangle. \tag{A.13}$$

Therefore the second term of (A.6) becomes:

$$\sum_l \sum_{S^{(l,n)}} \sum_{T^{(l,n)}} \delta^{(n)}_{P,Q,S,T} \left(1 + \sum_{t=1}^{n} \gamma^{(t)}_{P,Q,S,T} + \sum_{t=1}^{n} \chi^{(t)}_{P,Q,S,T}\right), \quad (A.14)$$

where we have defined:

$$\delta^{(n)}_{P,Q,S,T} \equiv \left(\bar{\tau}^{(n)} \tau^{(n)} - 1\right) C_S^{(k)\dagger} C_T^{(k)} \underset{\rightarrow}{\bar{\tau}}^{(n-1)} \underset{\rightarrow}{\tau}^{(n-1)}. \quad (A.15)$$

.

## A.4 Miss C → Decompose particle B partially

To analyze the first term of Eq. (A.6), transform the contributions from particle C back to a sum over paths, then collect the paths of particle B. Once this has been done, convert the paths of particle B to matrix notation. This might seem rather convoluted, since we are transforming back and forth between matrix and sum-over-path notations. However, the goal is quite clear; we are shifting the matrix notation of particle C onto particle B, by transferring the phases $\underset{\rightarrow}{\bar{\varepsilon}}^{(n)}$ and $\underset{\rightarrow}{\varepsilon}^{(n)}$.

$$\sum_k \sum_{M^{(k,n)}} \sum_{N^{(k,n)}} B_M^* B_N^* \underset{\rightarrow}{\bar{\eta}}^{(n)} \underset{\rightarrow}{\eta}^{(n)} \langle 0 | \mathbb{C}_{P,M}^{(n-1)\dagger} \mathbb{C}^{(n)\dagger} \mathbb{B} \mathbb{C}_M^{(n)\dagger} \mathbb{B} \mathbb{C}_N^{(n)} \mathbb{C}^{(n)} \mathbb{C}_{Q,N}^{(n-1)} | 0 \rangle$$

$$= \sum_k \sum_{M^{(k,n)}} \sum_{N^{(k,n)}} B_M^* B_N^* \underset{\rightarrow}{\bar{\eta}}^{(n)} \underset{\rightarrow}{\eta}^{(n)} \sum_l \left( \sum_{S^{(l,n)}} C_S^{(k)\dagger} \underset{\rightarrow}{\bar{\tau}}^{(n-1)} \underset{\rightarrow}{\bar{\varepsilon}}^{(n)} \right) \left( \sum_{T^{(l,n)}} C_T^{(k)} \underset{\rightarrow}{\tau}^{(n-1)} \underset{\rightarrow}{\varepsilon}^{(n)} \right)$$

$$= \sum_l \sum_{S^{(l,n)}} \sum_{T^{(l,n)}} \sum_k \left( \sum_{M^{(k,n)}} B_M^* \underset{\rightarrow}{\bar{\eta}}^{(n)} \underset{\rightarrow}{\bar{\varepsilon}}^{(n)} \right) \left( \sum_{N^{(k,n)}} B_M^* \underset{\rightarrow}{\eta}^{(n)} \underset{\rightarrow}{\varepsilon}^{(n)} \right) C_S^{(k)\dagger} C_T^{(k)} \underset{\rightarrow}{\bar{\tau}}^{(n-1)} \underset{\rightarrow}{\tau}^{(n-1)}$$

$$= \sum_l \sum_{S^{(l,n)}} \sum_{T^{(l,n)}} \sum_k \langle 0 | \mathbb{B}_{P,S}^\dagger | k \rangle \langle k | \mathbb{B}_{Q,T} | 0 \rangle C_S^{(k)\dagger} C_T^{(k)} \underset{\rightarrow}{\bar{\tau}}^{(n-1)} \underset{\rightarrow}{\tau}^{(n-1)}$$

$$= \sum_l \sum_{S^{(l,n)}} \sum_{T^{(l,n)}} \sum_k \langle 0 | \mathbb{B}_{P,S}^\dagger \mathbb{B}_{Q,T} | 0 \rangle C_S^{(k)\dagger} C_T^{(k)} \underset{\rightarrow}{\bar{\tau}}^{(n-1)} \underset{\rightarrow}{\tau}^{(n-1)}. \quad (A.16)$$

Now decompose $\langle 0 | \mathbb{B}_{P,S}^\dagger \mathbb{B}_{Q,T} | 0 \rangle$ a single time using the expansion:

$$\mathbb{AB}_P^{(n)\dagger} \mathbb{AB}_Q^{(n)} = \mathbb{I} + \sum_k \left( \bar{\eta}^{(n)} \eta^{(n)} - 1 \right) |k\rangle \langle k|. \quad (A.17)$$

Therefore:

$$\langle 0 | \mathbb{B}_{P,S}^\dagger \mathbb{B}_{Q,T} | 0 \rangle = \langle 0 | \mathbb{B}_{P,S}^{(n-1)\dagger} \mathbb{B}^{(n)\dagger} \mathbb{BC}_S^{(n)\dagger} \mathbb{AB}_P^{(n)\dagger} \mathbb{AB}_Q^{(n)} \mathbb{BC}_T^{(n)} \mathbb{B}^{(n)} \mathbb{B}_{Q,T}^{(n-1)} | 0 \rangle$$

$$= \langle 0 | \mathbb{B}_{P,S}^{(n-1)\dagger} \mathbb{B}^{(n)\dagger} \mathbb{BC}_S^{(n)\dagger} \mathbb{BC}_T^{(n)} \mathbb{B}^{(n)} \mathbb{B}_{Q,T}^{(n-1)} | 0 \rangle \quad (A.18)$$

$$+ \sum_k \left( \bar{\eta}^{(n)} \eta^{(n)} - 1 \right) \langle 0 | \mathbb{B}_{P,S}^{(n-1)\dagger} \mathbb{B}^{(n)\dagger} \mathbb{BC}_S^{(n)\dagger} | k \rangle \langle k | \mathbb{BC}_T^{(n)} \mathbb{B}^{(n)} \mathbb{B}_{Q,T}^{(n-1)} | 0 \rangle.$$

Consequently, the first term in the decomposition expressed by Eq. (A.6) is equal to:

$$\sum_l \sum_{S^{(l,n)}} \sum_{T^{(l,n)}} C_S^\dagger C_T \underset{\rightarrow}{\bar{\tau}}^{(n-1)} \underset{\rightarrow}{\tau}^{(n-1)} \left( \langle 0 | \mathbb{B}_{P,S}^{(n-1)\dagger} \mathbb{B}^{(n)\dagger} \mathbb{BC}_S^{(n)\dagger} \mathbb{BC}_T^{(n)} \mathbb{B}^{(n)} \mathbb{B}_{Q,T}^{(n-1)} | 0 \rangle \right) \quad (A.19)$$

$$+ \sum_l \sum_{S^{(l,n)}} \sum_{T^{(l,n)}} C_S^\dagger C_T \underset{\rightarrow}{\bar{\tau}}^{(n-1)} \underset{\rightarrow}{\tau}^{(n-1)} \sum_k \left( \bar{\eta}^{(n)} \eta^{(n)} - 1 \right) \langle 0 | \mathbb{B}_{P,S}^{(n-1)\dagger} \mathbb{B}^{(n)\dagger} \mathbb{BC}_S^{(n)\dagger} | k \rangle \langle k | \mathbb{BC}_T^{(n)} \mathbb{B}^{(n)} \mathbb{B}_{Q,T}^{(n-1)} | 0 \rangle.$$

Both of these terms need to be evaluated further.

## A.5 Hit B → Decompose C fully → End

Starting with the second term of Eq. (A.19), convert the contributions from particle B into a sum over paths, then collect the paths of particle C. Once this has been done, convert the paths of particle C to matrix notation. Notice again the transfer of the phases $\underrightarrow{\bar{\varepsilon}}^{(n)}$ and $\underrightarrow{\varepsilon}^{(n)}$.

$$\sum_{l} \sum_{S^{(l,n)}} \sum_{T^{(l,n)}} C_S^\dagger C_T \underrightarrow{\bar{\tau}}^{(n-1)} \underrightarrow{\tau}^{(n-1)} \sum_k \left( \bar{\eta}^{(n)} \eta^{(n)} - 1 \right) \langle 0 | \mathbb{B}_{P,S}^{(n-1)\dagger} \mathbb{B}^{(n)\dagger} \mathbb{B} \mathbb{C}_S^{(n)\dagger} | k \rangle \langle k | \mathbb{B} \mathbb{C}_T^{(n)} \mathbb{B}^{(n)} \mathbb{B}_{Q,T}^{(n-1)} | 0 \rangle$$

$$= \sum_{l} \sum_{S^{(l,n)}} \sum_{T^{(l,n)}} C_S^\dagger C_T \underrightarrow{\bar{\tau}}^{(n-1)} \underrightarrow{\tau}^{(n-1)} \sum_k \left( \bar{\eta}^{(n)} \eta^{(n)} - 1 \right) \left( \sum_{M^{(k,n)}} B_M^* \underrightarrow{\bar{\eta}}^{(n)} \underrightarrow{\bar{\varepsilon}}^{(n)} \right) \left( \sum_{N^{(k,n)}} B_M^* \underrightarrow{\eta}^{(n)} \underrightarrow{\varepsilon}^{(n)} \right)$$

$$= \sum_{k} \sum_{M^{(k,n)}} \sum_{N^{(k,n)}} \left( \bar{\eta}^{(n)} \eta^{(n)} - 1 \right) B_M^* B_N^* \underrightarrow{\bar{\eta}}^{(n)} \underrightarrow{\eta}^{(n)} \sum_{l} \left( \sum_{S^{(l,n)}} C_S^\dagger \underrightarrow{\bar{\tau}}^{(n-1)} \underrightarrow{\bar{\varepsilon}}^{(n)} \right) \left( \sum_{T^{(l,n)}} C_T \underrightarrow{\tau}^{(n-1)} \underrightarrow{\varepsilon}^{(n)} \right)$$

$$= \sum_{k} \sum_{M^{(k,n)}} \sum_{N^{(k,n)}} \left( \bar{\eta}^{(n)} \eta^{(n)} - 1 \right) B_M^* B_N^* \underrightarrow{\bar{\eta}}^{(n)} \underrightarrow{\eta}^{(n)} \sum_{l} \left( \langle 0 | \mathbb{C}_{P,M \nmid n}^{(n)\dagger} | l \rangle \right) \left( \langle l | \mathbb{C}_{Q,N \nmid n}^{(n)} | 0 \rangle \right)$$

$$= \sum_{k} \sum_{M^{(k,n)}} \sum_{N^{(k,n)}} \left( \bar{\eta}^{(n)} \eta^{(n)} - 1 \right) B_M^* B_N^* \underrightarrow{\bar{\eta}}^{(n)} \underrightarrow{\eta}^{(n)} \langle 0 | \mathbb{C}_{P,M \nmid n}^{(n)\dagger} \mathbb{C}_{Q,N \nmid n}^{(n)} | 0 \rangle. \tag{A.20}$$

The matrices $\mathbb{C}_{Q,N \nmid n}^{(n)}$ are the unitary matrices for particle C up to layer n of the circuit, given that there are no phases due to A-C interactions at layer n. The subscript $Q, N \nmid n$ distinguishes $\mathbb{C}_{Q,N \nmid n}^{(n)}$ from $\mathbb{C}_{Q,N}^{(n)}$. We have defined the matrix in this way because particle C has already been missed at layer n, which removed the effect of the A-C interaction. Now decompose $\langle 0 | \mathbb{C}_{P,M \nmid n}^{(n)\dagger} \mathbb{C}_{Q,N \nmid n}^{(n)} | 0 \rangle$ fully by inserting the expansions:

$$\mathbb{A} \mathbb{C}_P^{(t)\dagger} \mathbb{A} \mathbb{C}_Q^{(t)} = \sum_l \left[ \mathbb{I} + \left( \bar{\tau}^{(t)} \tau^{(t)} - 1 \right) | l \rangle \langle l | \right]. \tag{A.21}$$

$$\mathbb{B} \mathbb{C}_N^{(t)\dagger} \mathbb{B} \mathbb{C}_M^{(t)} = \sum_l \left[ \mathbb{I} + \left( \bar{\varepsilon}^{(t)} \varepsilon^{(t)} - 1 \right) | l \rangle \langle l | \right]. \tag{A.22}$$

until the $|l\rangle \langle l|$ part of the expansions are hit, at which point the decomposition stops. The result is:

$$\langle 0 | \mathbb{C}_{P,M \nmid n}^{(n)\dagger} \mathbb{C}_{Q,N \nmid n}^{(n)} | 0 \rangle = 1 + \sum_{t=1}^{n-1} \sum_s \left( \bar{\tau}^{(t)} \tau^{(t)} - 1 \right) \langle 0 | \mathbb{C}_{P,M}^{(t-1)\dagger} \mathbb{C}^{(t)\dagger} \mathbb{B} \mathbb{C}_M^{(t)\dagger} | l \rangle \langle l | \mathbb{B} \mathbb{C}_N^{(t)} \mathbb{C}^{(t)} \mathbb{C}_{Q,N}^{(t-1)} | 0 \rangle$$

$$+ \sum_{t=1}^n \sum_s \left( \bar{\varepsilon}^{(n)} \varepsilon^{(n)} - 1 \right) \langle 0 | \mathbb{C}_{P,M}^{(t-1)\dagger} \mathbb{C}^{(t)\dagger} | l \rangle \langle l | \mathbb{C}^{(t)} \mathbb{C}_{Q,N}^{(t-1)} | 0 \rangle$$

$$= 1 + \sum_{t=1}^{n-1} \gamma_{P,Q,M,N}^{(t)} + \sum_{t=1}^n \chi_{P,Q,M,N}^{(t)}. \tag{A.23}$$

We have defined:

$$\gamma_{P,Q,M,N}^{(t)} \equiv \sum_l \left( \bar{\tau}^{(t)} \tau^{(t)} - 1 \right) \langle 0 | \mathbb{C}_{P,M}^{(t-1)\dagger} \mathbb{C}^{(t)\dagger} \mathbb{B} \mathbb{C}_M^{(t)\dagger} | l \rangle \langle l | \mathbb{B} \mathbb{C}_N^{(t)} \mathbb{C}^{(t)} \mathbb{C}_{Q,N}^{(t-1)} | 0 \rangle. \tag{A.24}$$

$$\chi_{P,Q,M,N}^{(t)} \equiv \sum_l \left( \bar{\varepsilon}^{(t)} \varepsilon^{(t)} - 1 \right) \langle 0 | \mathbb{C}_{P,M}^{(t-1)\dagger} \mathbb{C}^{(t)\dagger} | l \rangle \langle l | \mathbb{C}^{(t)} \mathbb{C}_{Q,N}^{(t-1)} | 0 \rangle. \tag{A.25}$$

Due to Eq. (A.23) and Eq. (A.20) the second term of Eq. (A.19) becomes:

$$\sum_k \sum_{M^{(k,n)}} \sum_{N^{(k,n)}} \delta^{(n)}_{M,N,P,Q} \left(1 + \sum_{t=1}^{n-1} \gamma^{(t)}_{P,Q,M,N} + \sum_{t=1}^{n} \chi^{(t)}_{P,Q,M,N}\right), \quad (A.26)$$

where we have defined:

$$\delta^{(n)}_{M,N,P,Q} = \left(\bar{\eta}^{(n)} \eta^{(n)} - 1\right) B_M^* B_N^* \underset{\rightarrow}{\bar{\eta}}^{(n-1)} \underset{\rightarrow}{\eta}^{(n-1)}. \quad (A.27)$$

## A.6 Miss B → Remove B-C gate → Begin recursive sequence

If both particle C and B have been missed at layer n, then we need to remove the effect of the B-C interaction at this layer, which is encoded by the gate $\mathbb{BC}^{(n)}$. The fact that the effect of the B-C interaction can be removed from the marginals of particle A is physically evident from the no-signaling property. Mathematically, we can remove the $\mathbb{BC}^{(n)}$ gate in the following manner. The expression corresponding to a miss of both particle's B and C is:

$$\sum_l \sum_{S^{(l,n)}} \sum_{T^{(l,n)}} C_S^\dagger C_T \underset{\rightarrow}{\bar{\tau}}^{(n-1)} \underset{\rightarrow}{\tau}^{(n-1)} \left(\langle 0|\mathbb{B}^{(n-1)\dagger}_{P,S} \mathbb{B}^{(n)\dagger} \mathbb{BC}^{(n)\dagger}_S \mathbb{BC}^{(n)}_T \mathbb{B}^{(n)} \mathbb{B}^{(n-1)}_{Q,T}|0\rangle\right)$$

$$= \sum_l \sum_{S^{(l,n)}} \sum_{T^{(l,n)}} C_S^\dagger C_T \underset{\rightarrow}{\bar{\tau}}^{(n-1)} \underset{\rightarrow}{\tau}^{(n-1)} \sum_k \left(\sum_{M^{(k,n)}} B_M^\dagger \underset{\rightarrow}{\bar{\eta}}^{(n-1)} \underset{\rightarrow}{\bar{\varepsilon}}^{(n)}\right) \left(\sum_{N^{(k,n)}} B_N \underset{\rightarrow}{\eta}^{(n-1)} \underset{\rightarrow}{\varepsilon}^{(n)}\right)$$

$$= \sum_k \sum_l \sum_{M^{(k,n)}} \sum_{N^{(k,n)}} \sum_{S^{(l,n)}} B_M^\dagger B_N C_S^\dagger C_T \underset{\rightarrow}{\bar{\eta}}^{(n-1)} \underset{\rightarrow}{\eta}^{(n-1)} \underset{\rightarrow}{\bar{\tau}}^{(n-1)} \underset{\rightarrow}{\tau}^{(n-1)} \underset{\rightarrow}{\bar{\varepsilon}}^{(n)} \underset{\rightarrow}{\varepsilon}^{(n)} \quad (A.28)$$

$$= \sum_k \sum_l \langle 0|\mathbb{U}^{(n-1)\dagger}_P \left(\mathbb{A}^{(n)\dagger} \otimes \mathbb{B}^{(n)\dagger}\right) \mathbb{AB}^{(n)\dagger} |kl\rangle\langle kl|\mathbb{AB}^{(n)} \left(\mathbb{A}^{(n)} \otimes \mathbb{B}^{(n)}\right) \mathbb{U}^{(n-1)}_Q|0\rangle$$

$$= \langle 0|\mathbb{U}^{(n-1)\dagger}_P \left(\mathbb{A}^{(n)\dagger} \otimes \mathbb{B}^{(n)\dagger}\right) \mathbb{AB}^{(n)\dagger} \left[\sum_k |k\rangle\langle k| \otimes \sum_l |l\rangle\langle l|\right] \mathbb{AB}^{(n)} \left(\mathbb{A}^{(n)} \otimes \mathbb{B}^{(n)}\right) \mathbb{U}^{(n-1)}_Q|0\rangle$$

$$= \langle 0|\mathbb{U}^{(n-1)\dagger}_P \left(\mathbb{A}^{(n)\dagger} \otimes \mathbb{B}^{(n)\dagger}\right) \mathbb{AB}^{(n)\dagger} \mathbb{AB}^{(n)} \left(\mathbb{A}^{(n)} \otimes \mathbb{B}^{(n)}\right) \mathbb{U}^{(n-1)}_Q|0\rangle \quad (A.29)$$

$$= \langle 0|\mathbb{U}^{(n-1)\dagger}_P \mathbb{U}^{(n-1)}_Q|0\rangle. \quad (A.30)$$

We have converted the path sum in Eq. (A.28) to matrix notation by studying the phases present. The only controlled phase gate present at layer n is the gate $\mathbb{AB}^{(n)}$, which arises as a consequence of the of phases $\underset{\rightarrow}{\varepsilon}^{(n)}$ and $\underset{\rightarrow}{\bar{\varepsilon}}^{(n)}$.

After removing the $\mathbb{AB}^{(n)}$ the final result $\langle 0|\mathbb{U}^{(n-1)\dagger}\mathbb{U}^{(n-1)}|0\rangle$ is of the same general form as $\langle 0|\mathbb{U}^{(n)\dagger}\mathbb{U}^{(n)}|0\rangle$ but for one layer earlier in the circuit. Therefore we can recursively apply the decomposition procedure. Explicitly, for the first step we have:

$$\langle 0|\mathbb{U}^{(n)\dagger}\mathbb{U}^{(n)}|0\rangle = \sum_k \sum_{M^{(k,n)}} \sum_{N^{(k,n)}} \delta^{(n)}_{M,N,P,Q} \left(1 + \sum_{t=1}^{n-1} \gamma^{(t)}_{P,Q,M,N} + \sum_{t=1}^{n} \chi^{(t)}_{P,Q,M,N}\right) \quad (A.31)$$

$$+ \sum_l \sum_{S^{(l,n)}} \sum_{T^{(l,n)}} \delta^{(n)}_{P,Q,S,T} \left(1 + \sum_{t=1}^{n} \gamma^{(t)}_{P,Q,S,T} + \sum_{t=1}^{n} \chi^{(t)}_{P,Q,S,T}\right)$$

$$+ \langle 0|\mathbb{U}^{(n-1)\dagger}\mathbb{U}^{(n-1)}|0\rangle.$$

And for the second step:

$$\langle 0|\mathbb{U}^{(n-1)\dagger}\mathbb{U}^{(n-1)}|0\rangle = \sum_k \sum_{M^{(k,n-1)}} \sum_{N^{(k,n-1)}} \delta^{(n-1)}_{M,N,P,Q} \left(1 + \sum_{t=1}^{n-2} \gamma^{(t)}_{P,Q,M,N} + \sum_{t=1}^{n-1} \chi^{(t)}_{P,Q,M,N}\right) \quad (A.32)$$

$$+ \sum_l \sum_{S^{(l,n-1)}} \sum_{T^{(l,n-1)}} \delta^{(n-1)}_{P,Q,S,T} \left(1 + \sum_{t=1}^{n-1} \gamma^{(t)}_{P,Q,S,T} + \sum_{t=1}^{n-1} \chi^{(t)}_{P,Q,S,T}\right)$$

$$+ \langle 0|\mathbb{U}^{(n-2)\dagger}\mathbb{U}^{(n-2)}|0\rangle.$$

And so forth. The last decomposition produces $\langle 0|\mathbb{U}^{(n-n)\dagger}\mathbb{U}^{(n-n)}|0\rangle = 1$. Therefore:

$$\langle 0|\mathbb{U}^{(n)\dagger}\mathbb{U}^{(n)}|0\rangle = 1 + \sum_{r=1}^{n} \left[\sum_k \sum_{M^{(k,r)}} \sum_{N^{(k,r)}} \delta^{(r)}_{M,N,P,Q} \left(1 + \sum_{t=1}^{r-1} \gamma^{(t)}_{P,Q,M,N} + \sum_{t=1}^{r} \chi^{(t)}_{P,Q,M,N}\right)\right] \quad (A.33)$$

$$+ \sum_{r=1}^{n} \left[\sum_l \sum_{S^{(l,r)}} \sum_{T^{(l,r)}} \delta^{(r)}_{P,Q,S,T} \left(1 + \sum_{t=1}^{r} \gamma^{(t)}_{P,Q,S,T} + \sum_{t=1}^{r} \chi^{(t)}_{P,Q,S,T}\right)\right].$$

# B Reduced density matrix ontology

In this section, we will briefly discuss whether reduced density matrices provide a local description of quantum subsystems. Suppose we have two particles A and B which are entangled in spatial mode by a quantum circuit. The density matrix for the total system at layer $t$ of the circuit is $\rho^{(t)}_{AB} = |\Psi(t)\rangle\langle\Psi(t)|$. The reduced density matrix of particle A at layer $t$ is given by a partial trace over the density matrix $\rho^{(t)}_{AB}$:

$$\rho^{(t)}_A = Tr_B\left[\rho^{(t)}_{AB}\right]$$
$$= Tr_B\left[\mathbb{AB}^{(t)}\mathbb{A}^{(t)} \otimes \mathbb{B}^{(t)} \rho^{(t-1)}_{AB} \mathbb{B}^{(t)\dagger} \otimes \mathbb{A}^{(t)\dagger}\mathbb{AB}^{(t)\dagger}\right], \quad (B.1)$$

where we have expanded out the first layer of the reduced density matrix using:

$$|\Psi(t)\rangle = \mathbb{AB}^{(t)}\mathbb{A}^{(t)} \otimes \mathbb{B}^{(t)}|\Psi(t-1)\rangle, \quad (B.2)$$

where $\mathbb{AB}^{(t)}$ is a controlled phase gate applied at layer $t$ of the circuit, and $\mathbb{A}^{(t)}$, $\mathbb{B}^{(t)}$ are the single particle gates applied at layer $t$. By explicitly writing the partial trace, the expression for the reduced density matrix becomes:

$$\rho^{(t)}_A = \langle 0|_B \left(\mathbb{A}^{(t)} \otimes \mathbb{B}^{(t)}\right) \rho^{(t-1)}_{AB} \left(\mathbb{B}^{(t)\dagger} \otimes \mathbb{A}^{(t)\dagger}\right) |0\rangle_B \quad (B.3)$$
$$+ \langle 1|_B \left(\mathbb{Z}_\theta \mathbb{A}^{(t)} \otimes \mathbb{B}^{(t)}\right) \rho^{(t-1)}_{AB} \left(\mathbb{B}^{(t)\dagger}\right) \otimes \mathbb{A}^{(t)\dagger}\mathbb{Z}^\dagger_\theta|1\rangle_B,$$

where we have used $\mathbb{AB}^{(t)}(\mathbb{I} \otimes |0\rangle_B) = \mathbb{I} \otimes |0\rangle_B$ and $\mathbb{AB}^{(t)}(\mathbb{I} \otimes |1\rangle_B) = \mathbb{Z}_\theta \otimes |1\rangle_B$, assuming:

$$\mathbb{AB}^{(t)} = \mathbb{I} \otimes |0\rangle_B\langle 0|_B + \mathbb{Z}_\theta \otimes |1\rangle_B\langle 1|_B, \quad (B.4)$$

for a simple example, where the gate $\mathbb{Z}_\theta$ is taken to be $\mathbb{Z}_\theta = \text{diag}\{1, e^{i\theta}\}$. This gate describes the phases applied to particle A given that particle B is in the mode $|1\rangle_B$ at layer $t$. The two pieces of Eq. (B.3) are non-local, since they describe the probabilities for wavefunction collapse when measured.

However, there is a transformation which can be applied that makes the non-locality less evident. First, use the cyclic property of the partial trace to convert this expression to:

$$\begin{aligned}\rho_A^{(t)} =& Tr_B\left[|0\rangle_B\langle 0|_B \left(\mathbb{A}^{(t)}\otimes\mathbb{B}^{(t)}\right)\rho_{AB}^{(t-1)}\left(\mathbb{B}^{(t)\dagger}\otimes\mathbb{A}^{(t)\dagger}\right)\right]\\ &+\langle 1|_B\left(\mathbb{Z}_\theta\mathbb{A}^{(t)}\otimes\mathbb{B}^{(t)}\right)\rho_{AB}^{(t-1)}\left(\mathbb{B}^{(t)\dagger}\right)\otimes\mathbb{A}^{(t)\dagger}\mathbb{Z}_\theta^\dagger|1\rangle_B,\end{aligned} \quad\text{(B.5)}$$

Now substitute $|0\rangle_B\langle 0|_B = \mathbb{I}_B - |1\rangle_B\langle 1|_B$ into the first term of Eq. (B.5).

$$\begin{aligned}\rho_A^{(t)} =& Tr_B\left[(\mathbb{I}_B - |1\rangle_B\langle 1|_B)\left(\mathbb{A}^{(t)}\otimes\mathbb{B}^{(t)}\right)\rho_{AB}^{(t-1)}\left(\mathbb{B}^{(t)\dagger}\otimes\mathbb{A}^{(t)\dagger}\right)\right]\\ &+\langle 1|_B\left(\mathbb{Z}_\theta\mathbb{A}^{(t)}\otimes\mathbb{B}^{(t)}\right)\rho_{AB}^{(t-1)}\left(\mathbb{B}^{(t)\dagger}\right)\otimes\mathbb{A}^{(t)\dagger}\mathbb{Z}_\theta^\dagger|1\rangle_B\\ =& Tr_B\left[\mathbb{A}^{(t)}\otimes\mathbb{B}^{(t)}\rho_{AB}^{(t-1)}\left(\mathbb{B}^{(t)\dagger}\otimes\mathbb{A}^{(t)\dagger}\right)\right] - \langle 1|_B\left(\mathbb{A}^{(t)}\otimes\mathbb{B}^{(t)}\right)\rho_{AB}^{(t-1)}\left(\mathbb{B}^{(t)\dagger}\right)\otimes\mathbb{A}^{(t)\dagger}|1\rangle_B\\ &+\langle 1|_B\left(\mathbb{Z}_\theta\mathbb{A}^{(t)}\otimes\mathbb{B}^{(t)}\right)\rho_{AB}^{(t-1)}\left(\mathbb{B}^{(t)\dagger}\right)\otimes\mathbb{A}^{(t)\dagger}\mathbb{Z}_\theta^\dagger|1\rangle_B\end{aligned} \quad\text{(B.6)}$$

By using the cyclic property of the partial trace on the first term of (B.6), the reduced density matrix for particle A becomes:

$$\begin{aligned}\rho_A^{(t)} =& \mathbb{A}^{(t)}\rho_A^{(t-1)}\mathbb{A}^{(t)\dagger}\\ &-\langle 1|_B\mathbb{A}^{(t)}\otimes\mathbb{B}^{(t)}\rho_{AB}^{(t-1)}\mathbb{B}^{(t)\dagger}\mathbb{A}^{(t)\dagger}|1\rangle_B + \langle 1|_B\mathbb{Z}_\theta\mathbb{A}^{(t)}\otimes\mathbb{B}^{(t)}\rho_{AB}^{(t-1)}\mathbb{B}^{(t)\dagger}\otimes\mathbb{A}^{(t)\dagger}\mathbb{Z}_\theta^\dagger|1\rangle_B\\ \equiv& \rho_{A|\text{miss}}^{(t)} + \rho_{A|\text{hit}}^{(t)}\end{aligned} \quad\text{(B.7)}$$

where $\rho_{A|\text{miss}}^{(t)}$ and $\rho_{A|\text{hit}}^{(t)}$ are defined as:

$$\rho_{A|\text{miss}}^{(t)} \equiv \mathbb{A}^{(t)}\rho_A^{(t-1)}\mathbb{A}^{(t)\dagger}. \qquad \begin{aligned}\rho_{A|\text{hit}}^{(t)} \equiv& -\langle 1|_B\mathbb{A}^{(t)}\otimes\mathbb{B}^{(t)}\rho_{AB}^{(t-1)}\mathbb{B}^{(t)\dagger}\otimes\mathbb{A}^{(t)\dagger}|1\rangle_B\\ &+\langle 1|_B\mathbb{Z}_\theta\mathbb{A}^{(t)}\otimes\mathbb{B}^{(t)}\rho_{AB}^{(t-1)}\mathbb{B}^{(t)\dagger}\otimes\mathbb{A}^{(t)\dagger}\mathbb{Z}_\theta^\dagger|1\rangle_B.\end{aligned}$$

The term $\rho_{A|\text{miss}}^{(t)}$ is essentially the reduced density matrix at layer $t$ given no interaction occurred. However, it would be misleading to regard $\rho_{A|\text{hit}}^{(t)}$ as a density matrix, since it does not independently give a probability when measured by particle A. It can perhaps be understood as a hidden variable for the subsystem.

Now suppose that particle A knows the previous reduced density matrix $\rho_A^{(t-1)}$. The question is whether particle A can use this knowledge to construct $\rho_A^{(t)}$ after the interaction at layer $t$ has occurred. Clearly particle A can determine $\rho_{A|\text{miss}}^{(t)} = \mathbb{A}^{(t)}\rho_A^{(t-1)}\mathbb{A}^{(t)\dagger}$ from $\rho_A^{(t-1)}$. So we just need to determine whether particle A can construct $\rho_{A|\text{hit}}^{(t)}$. By inserting $\mathbb{I} = |0\rangle_A\langle 0|_A + |1\rangle_A\langle 1|_A$ into the two terms of $\rho_{A|\text{hit}}^{(t)}$, we get:

$$\begin{aligned}\langle 1|_B\mathbb{A}^{(t)}\otimes\mathbb{B}^{(t)}\rho_{AB}^{(t-1)}\mathbb{B}^{(t)\dagger}\otimes\mathbb{A}^{(t)\dagger}|1\rangle_B =& |0\rangle_A\langle 01|\mathbb{A}^{(t)}\otimes\mathbb{B}^{(t)}\rho_{AB}^{(t-1)}\mathbb{B}^{(t)\dagger}\otimes\mathbb{A}^{(t)\dagger}|10\rangle\langle 0|_A\\ &+|0\rangle_A\langle 01|\mathbb{A}^{(t)}\otimes\mathbb{B}^{(t)}\rho_{AB}^{(t-1)}\mathbb{B}^{(t)\dagger}\otimes\mathbb{A}^{(t)\dagger}|11\rangle\langle 1|_A\\ &+|1\rangle_A\langle 11|\mathbb{A}^{(t)}\otimes\mathbb{B}^{(t)}\rho_{AB}^{(t-1)}\mathbb{B}^{(t)\dagger}\otimes\mathbb{A}^{(t)\dagger}|10\rangle\langle 0|_A\\ &+|1\rangle_A\langle 11|\mathbb{A}^{(t)}\otimes\mathbb{B}^{(t)}\rho_{AB}^{(t-1)}\mathbb{B}^{(t)\dagger}\otimes\mathbb{A}^{(t)\dagger}|11\rangle\langle 1|_A.\end{aligned} \quad\text{(B.8)}$$

$$\langle 1|_B \mathbb{A}^{(t)} \otimes \mathbb{B}^{(t)} \rho_{AB}^{(t-1)} \mathbb{B}^{(t)\dagger} \otimes \mathbb{A}^{(t)\dagger} |1\rangle_B = |0\rangle_A \langle 01| \mathbb{A}^{(t)} \otimes \mathbb{B}^{(t)} \rho_{AB}^{(t-1)} \mathbb{B}^{(t)\dagger} \otimes \mathbb{A}^{(t)\dagger} |10\rangle \langle 0|_A \quad \text{(B.9)}$$
$$+ e^{-i\theta} |0\rangle_A \langle 01| \mathbb{A}^{(t)} \otimes \mathbb{B}^{(t)} \rho_{AB}^{(t-1)} \mathbb{B}^{(t)\dagger} \otimes \mathbb{A}^{(t)\dagger} |11\rangle \langle 1|_A$$
$$+ e^{i\theta} |1\rangle_A \langle 11| \mathbb{A}^{(t)} \otimes \mathbb{B}^{(t)} \rho_{AB}^{(t-1)} \mathbb{B}^{(t)\dagger} \otimes \mathbb{A}^{(t)\dagger} |10\rangle \langle 0|_A$$
$$+ |1\rangle_A \langle 11| \mathbb{A}^{(t)} \otimes \mathbb{B}^{(t)} \rho_{AB}^{(t-1)} \mathbb{B}^{(t)\dagger} \otimes \mathbb{A}^{(t)\dagger} |11\rangle \langle 1|_A.$$

Subtracting the first from the second causes the diagonal elements of $\rho_{A|\text{hit}}^{(t)}$ to cancel, leaving the off-diagonal elements:

$$\rho_{A|\text{hit}}^{(t)} = \left( e^{-i\theta} - 1 \right) |0\rangle_A \langle 01| \mathbb{A}^{(t)} \otimes \mathbb{B}^{(t)} \rho_{AB}^{(t-1)} \mathbb{B}^{(t)\dagger} \otimes \mathbb{A}^{(t)\dagger} |11\rangle \langle 1|_A$$
$$+ \left( e^{i\theta} - 1 \right) |1\rangle_A \langle 11| \mathbb{A}^{(t)} \otimes \mathbb{B}^{(t)} \rho_{AB}^{(t-1)} \mathbb{B}^{(t)\dagger} \otimes \mathbb{A}^{(t)\dagger} |10\rangle \langle 0|_A$$
$$= |0\rangle_A \langle 1|_A \left[ \left( e^{-i\theta} - 1 \right) \langle 01| \mathbb{A}^{(t)} \otimes \mathbb{B}^{(t)} \rho_{AB}^{(t-1)} \mathbb{B}^{(t)\dagger} \otimes \mathbb{A}^{(t)\dagger} |11\rangle \right] \quad \text{(B.10)}$$
$$+ |1\rangle_A \langle 0|_A \left[ \left( e^{i\theta} - 1 \right) \langle 11| \mathbb{A}^{(t)} \otimes \mathbb{B}^{(t)} \rho_{AB}^{(t-1)} \mathbb{B}^{(t)\dagger} \otimes \mathbb{A}^{(t)\dagger} |10\rangle \right].$$

Since particle A is able to sample the phase $e^{i\theta}$, the only remaining piece of information required to be sampled is the complex number $\langle 01| \mathbb{A}^{(t)} \otimes \mathbb{B}^{(t)} \rho_{AB}^{(t-1)} \mathbb{B}^{(t)\dagger} \mathbb{A}^{(t)\dagger} |11\rangle$. But analyzing this expression reveals a problem. We have $\rho_{AB}^{(t-1)} = |\Psi(t-1)\rangle \langle \Psi(t-1)|$, therefore:

$$\langle 01| \mathbb{A}^{(t)} \otimes \mathbb{B}^{(t)} \rho_{AB}^{(t-1)} \mathbb{B}^{(t)\dagger} \otimes \mathbb{A}^{(t)\dagger} |11\rangle = \langle 01| \mathbb{A}^{(t)} \otimes \mathbb{B}^{(t)} |\Psi(t-1)\rangle \langle \Psi(t-1)| \mathbb{B}^{(t)\dagger} \otimes \mathbb{A}^{(t)\dagger} |11\rangle$$

Suppose the wavefunction $|\Psi(t-1)\rangle$ at layer $t-1$ is non-separable. In this case, the term $\langle \Psi(t-1)| \mathbb{B}^{(t)\dagger} \otimes \mathbb{A}^{(t)\dagger} |11\rangle$ can be determined locally. But the expression $\langle 01| \mathbb{A}^{(t)} \otimes \mathbb{B}^{(t)} |\Psi(t-1)\rangle$ describes the probability amplitude for wavefunction collapse to the state $|0\rangle_A \otimes |1\rangle_B$. Particle A cannot sample this amplitude at the interaction location, since mode $|0\rangle_A$ is not undergoing the interaction. Consequently, $\rho_{A|\text{hit}}^{(t)}$ cannot by constructed at the interaction location.

To restore a reduced density matrix ontology to the subsystem, it is necessary to build a hidden variable theory which allows $\rho_{A|\text{hit}}^{(t)}$ to be constructed later in the circuit. This can be achieved by splitting the expression $\langle 01| \mathbb{A}^{(t)} \otimes \mathbb{B}^{(t)} |\Psi(t-1)\rangle$ into a sum over paths:

$$\langle 01| \mathbb{A}^{(t)} \otimes \mathbb{B}^{(t)} |\Psi(t-1)\rangle = \sum_{P^{(0,t)}} A_P \langle 1| \mathbb{B}_P^{(t-1)} |0\rangle$$

Particle A can sample the amplitudes $\langle 1| \mathbb{B}_P^{(t)} |0\rangle$ during the interaction, and later learn the amplitudes $A_P$ to construct $\rho_{A|\text{hit}}^{(t)}$. But a sum over paths is eventually required in some capacity to establish a local reduced density matrix ontology. This issue becomes even more pronounced in the three-particle case and beyond.